\documentclass[iop,appendixfloats]{emulateapj}
\usepackage{natbib}
\usepackage{graphicx}
\usepackage{epstopdf}
\usepackage{lineno}
\usepackage{footnote}
\usepackage{epsfig}
\usepackage{graphics}
\usepackage{amsmath}
\usepackage{amssymb}
\usepackage{xcolor}
\usepackage{tikz}

\usepackage{lineno}

\shorttitle{Search for high-redshift blazars with \textit{Fermi}/LAT}
\shortauthors{Kreter et al.}

\usepackage[utf8]{inputenc}
\usepackage[T1]{fontenc}
\usepackage{hyperref}

\begin{document}

\title{Search for high-redshift blazars with \textit{Fermi}/LAT}

\author{M. Kreter$^1$, A. Gokus$^{2,3}$, F. Krau\ss{}$^4$, M. Kadler$^2$, R. Ojha$^5$, S. Buson$^2$, J. Wilms$^3$\\ and M. B\"{o}ttcher$^1$}
\affil{$^1$Centre for Space Research, North-West University, Private Bag X6001, Potchefstroom 2520, South Africa}
\affil{$^2$Lehrstuhl f\"{u}r Astronomie, Universit\"{a}t W\"{u}rzburg, Emil-Fischer-Strasse 31, 97074 W\"{u}rzburg, Germany}
\affil{$^3$Dr. Remeis Sternwarte \& ECAP, Universit\"at Erlangen-N\"urnberg, Sternwartstrasse 7, 96049 Bamberg, Germany}
\affil{$^4$Department of Astronomy \& Astrophysics, Pennsylvania State University, University Park, PA, 16802, United States}
\affil{$^5$NASA Goddard Space Flight Center, Greenbelt, MD 20771, USA}
\email{michael@kreter.org}

\begin{abstract}
High-\textit{z} blazars (z\,$\geq 2.5$) are the most powerful class of persistent $\gamma$-ray sources in the Universe.
These objects possess the highest jet powers and luminosities and have black hole masses often in excess of $10^9$ solar masses. In addition, high-$z$ blazars are important cosmological probes and serve as test objects for blazar evolution models. Due to their large distance, their high-energy emission typically peaks below the GeV range, which makes them difficult to study with \textit{Fermi}/LAT. Therefore, only the very brightest objects are detectable and, to date, only a small number of high-z blazars have been detected with \textit{Fermi}/LAT.\\
In this work, we studied the monthly binned long-term $\gamma$-ray emission of a sample of 176 radio and optically detected blazars that have not been reported as known $\gamma$-ray sources in the 3FGL catalog. \\ 
In order to account for false-positive detections, we calculated monthly \textit{Fermi}/LAT light curves for a large sample of blank sky positions and derived the number of random fluctuations that we expect at various test statistic (TS) levels. For a given blazar, a detection of TS > 9 in at least one month is expected $\sim 15\%$ of the time. Although this rate is too high to secure detection of an individual source, half of our sample shows such single-month $\gamma$-ray activity, indicating a population of high-energy blazars at distances of up to z=5.2. Multiple TS > 9 monthly detections are unlikely to happen by chance, and we have detected several individual new sources in this way, including the most distant $\gamma$-ray blazar, BZQ\,J1430+4204 (z = 4.72). Finally, two new $\gamma$-ray blazars at redshifts of z = 3.63 and z = 3.11 are unambiguously detected via very significant (TS > 25) flares in individual monthly time bins.
\end{abstract}

\section{Introduction}
\hspace{-13pt}
The most luminous sources across the entire electromagnetic spectrum are active galactic nuclei (AGN). A fraction of AGN exhibit a collimated outflow of material from the central region, a jet. Depending on the orientation of an AGN on the night sky, we see its jets from different angles. An AGN with a jet pointing towards Earth is called a blazar and, due to the relativistic speed of the material in the jet, its emission appears brighter and more variable compared to other AGN.
Due to the extreme luminosities of blazars, it is possible to observe them even at large redshifts.
Studies by \citet{Ajello_2009, Toda_2020} revealed that the density of the most powerful blazars appears to peak at z $\sim$ 4.
These high-\textit{z} blazars seem to harbor supermassive black holes with masses often in excess of $10^9\,M_{\odot}$ \citep{Bloemen_1995,Ghisellini_2010b}, which means they have larger masses than most blazars of much lower redshift.
The presence of such massive black holes in the early Universe implies that they must have grown very quickly. 
Observations show that high-\textit{z} blazars accrete matter at $\sim 10$\,\%  of the Eddington rate \citep{Ghisellini_2010b}, indicating that they have already reached the end of the blazar formation process even at such a high redshift.  \citep{Ghisellini_2013}.

\begin{figure}[pt]
    \centering
  \centering
  \includegraphics[width=0.99\linewidth]{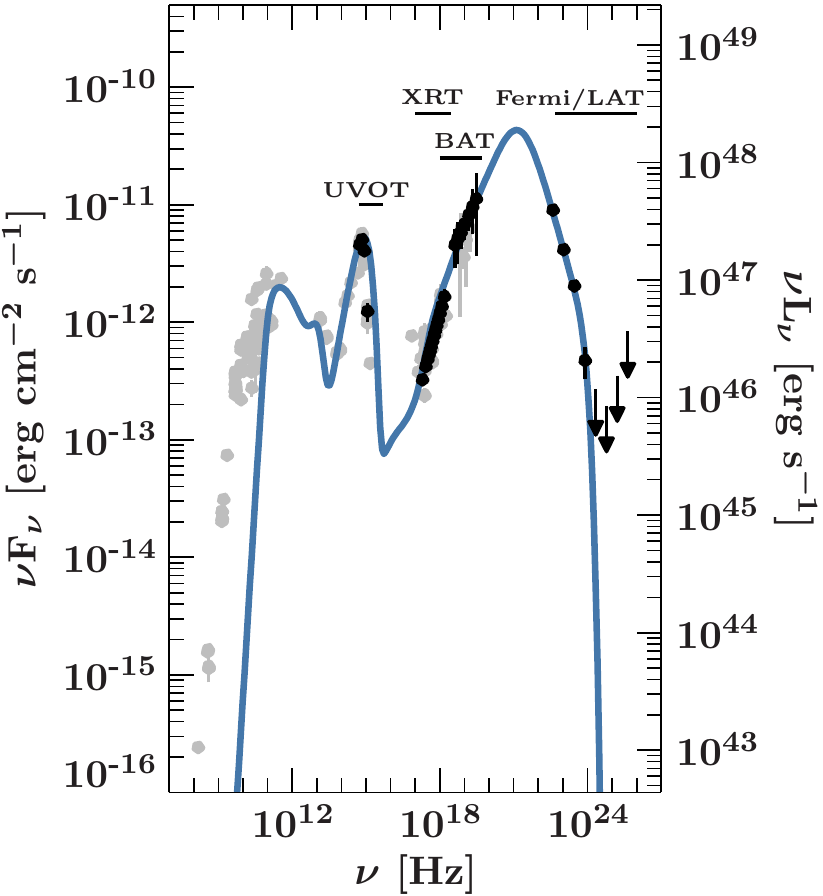} 
     \caption{Example spectral energy distribution of the high-$z$ blazar  DA\,193 (z = 2.365). These sources are typically weak at $\gamma$-ray energies, but show strong hard X-ray emission. Data adopted from \citet{Paliya_2019}}
   \label{SED_plot}
   \end{figure}

Their broadband emission is dominated by relativistic jets and shows the typical double-hump structure of blazars (see Fig.\,\ref{SED_plot}), but with the difference that the high-energy peak rises above the synchrotron peak by one order of magnitude or more in power \citep{Celotti_2008}.
Most of the known high-\textit{z} blazars have been detected in radio and optical surveys \citep{Condon_1998, Abolfathi_2018}.
While blazars, in general, dominate the $\gamma$-ray sky \citep{4FGL}, the Large Area Telescope (LAT) onboard the \textit{Fermi} satellite has only detected $\sim 12$ blazars at large ($z\geq2.5$) redshifts \citep{4FGL,Acero_2015}.
This discrepancy is caused by 
absorption on the extragalactic background light (EBL), which leads to an efficient degrading of high-energy $\gamma$ rays above $\sim 1$\,GeV \citep{Ackermann_2017,Hess_2013}.
This causes a suppression of the observed $\gamma$-ray emission from distant blazars.
Along with absorption on the EBL, galaxy evolution could also explain the low number of high-\textit{z} blazars observed by \textit{Fermi}/LAT \citep{Ajello_2014}. 

So far, the most distant known $\gamma$-ray emitting blazar is B3\,1428+422 with a redshift of $z = 4.72$ \citep{Liao_2018}. Many of the other high-$z$ blazars observed by \textit{Fermi}/LAT were only detectable when in bright flaring states \citep{Paliya_2019}. However, there is a lack of simultaneous broadband data, which is necessary for understanding their multiwavelength flaring behavior.
Currently, the only high-$z$ blazars observed at multiple frequencies during a flare are TXS\,0536+145 \citep[z = 2.69;][]{Orienti_2014} and CGRaBS\,J0733+0456\citep[z =3.01;][]{Liao_2019}. Their broadband emission is well described by a leptonic model, similar to models for blazars in the local Universe.
\cite{Paliya_2016,Paliya_2020,Berg_2019,Marcotulli_2020} studied spectral energy distributions (SEDs) of high-$z$ blazars, by using quasi-simultaneous datasets. All datasets could be well described by a simple one-zone leptonic model plus thermal emission from the accretion disk, with the emission from the high-energy hump originating from external Compton scattering.

Continuous monitoring of a chosen sample of sources in all available energy bands has revealed a correlation between the radio and $\gamma$-ray emission of blazars \citep[e.g.,][]{radio_gamma_corr, radio_gamma_corr2}, which is expected to also be observed for high-$z$ blazars. Their high jet power and possibly different evolutionary state makes them particularly interesting targets, but so far no study on the correlation between radio and $\gamma$-ray emission is available for the most distant blazars.
In the keV energy regime, the soft X-ray spectrum was observed to be flatter than expected for several high-$z$ sources \citep{flattening_xray}, which was first proposed to be due to intrinsic absorption of those sources \citep[e.g.,][]{intrinsic_absorption, intrinsic_absorption2}. However, their optical/UV emission did not show this
absorption, which created speculation about a so-called 'warm absorber', for which the absorbing material shows a huge gas-to-dust ratio and only affects the X-ray radiation \citep{warm_absorber}.
Another possibility is an intrinsic curvature within the jet emission, which does not require a strong absorber to explain the observed soft X-ray deficit \citep[e.g.,][]{Paliya_2016, curved_EC_emission2, curved_EC_emission}.

The number of known $\gamma$-ray emitting high-$z$ blazars was recently increased by \cite{Ackermann_2017}. These distant sources were detected by \textit{Fermi}/LAT due to an extended energy range towards lower energies and the employment of energy dispersion in the gamma-ray analysis, which allowed for the identification of MeV-dominated blazars.

In this work, we introduce a new method to search for undetected high-$z$ $\gamma$-ray emitting blazars that are too faint to be detected significantly on the long-term periods typically considered for catalogs, but can show significant emission during shorter ($\sim$ monthly) periods. This paper is structured in the following way:
In Sec.\,\ref{Sample}, we introduce the selection criteria for the considered sample of radio and optically detected blazars. Sec.\,\ref{LAT_analysis} describes the details of the \textit{Fermi}/LAT analysis performed. In Sec.\,\ref{Background determination} we discuss the approach employed to measure background fluctuations in the $\gamma$-ray sky and introduce statistical ways to calculate the probability of multiple low-significance\footnote{TS\,$< 25$ monthly intervals.} $\gamma$-ray detections from the same blazar. Sec.\,\ref{Individual Intervalls} focuses on
$\gamma$-ray detections from the considered blazar sample at various test statistic (TS) levels The test statistic is defined as a likelihood ratio test according to $ TS = -2\log (L/Lo)$.
In Sec.\,\ref{Discussion} we compare our work to the first catalog of transient $\gamma$-ray sources observed by \textit{Fermi}/LAT (1FLT) \citep{Mereu_inprep} and discuss the limitations and statistical uncertainties of the method used. Furthermore, we present SEDs for the most significantly detected high-\textit{z} blazars and conclude in Sec.\,\ref{Conclusion}.

\section{Method}
\subsection{Detection Strategy and Sample Selection} \label{Sample}
\begin{figure}[pt]
        \centering
   \includegraphics[width=0.99\linewidth]{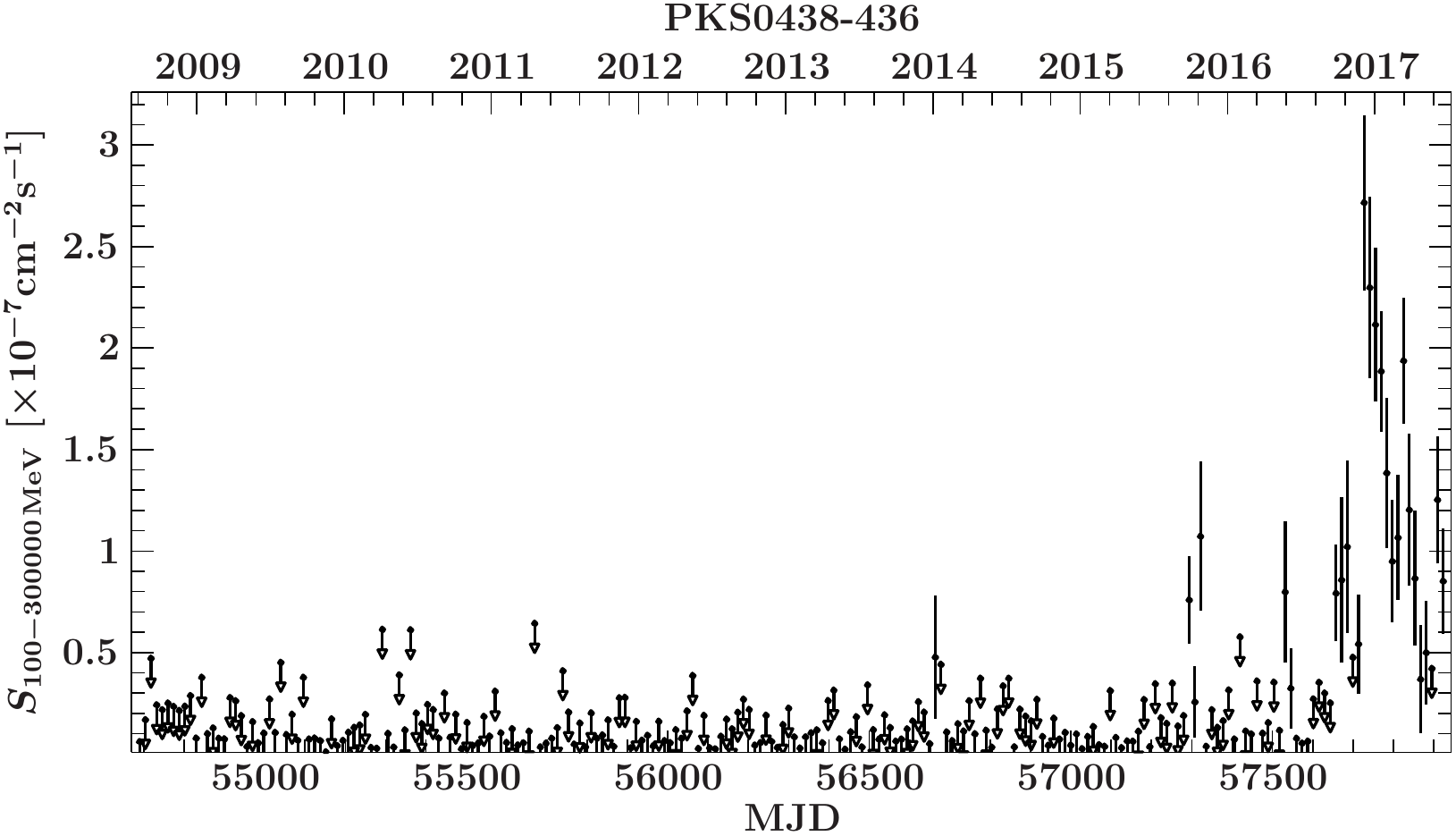}
     \caption{
  \textit{Fermi}/LAT $\gamma$-ray light curve of PKS\,0438$-$43 over about ten years of continuous observations. The light curve is shown for the energy range 100\,MeV to 300\,GeV.
  This source was detected in December 2016 on daily time scales during a bright flare.}
     \label{PKS0438LC}
\end{figure}

\hspace{-14.5pt}
Flaring activity from previously undetected high-$z$  $\gamma$-ray blazars has been discovered several times by \textit{Fermi}/LAT \citep{Cheung_2016,Cheung_2017,Angioni_2018}.~Such periods of enhanced activity typically only last for several days, leading to a significant detection (TS $\geq 25$) due to a drastic background reduction compared to the long-term averaged fluxes used in catalogs.
 Figure\,\ref{PKS0438LC} shows a ten-year \textit{Fermi}/LAT, $\gamma$-ray light curve for the high-$z$ blazar PKS\,0438$-$43 (z = 2.83) in the energy range 100\,MeV to 300\,GeV. 
 This source was  not included in the third \textit{Fermi}/LAT Source catalog and was detected at GeV energies only due to its bright flare in 2016\footnote{PKS\,0438$-$43 is now included in the Fourth \textit{Fermi}/LAT Source catalog (4FGL).} \citep{Cheung_2016}. \\
  
\hspace{-14pt}
In order to search for such flaring activity, we perform a systematic search for high-$z$ blazars that are not present in any \textit{Fermi}/LAT catalogs\footnote{Since this project started before the release of the 4FGL catalog, we used the previous 3FGL \textit{Fermi}/LAT source catalog. Some of the blazars that we found as part of our study were also found by the 4FGL pipeline.},
by targeting a sample of 176 radio and optically detected blazars with a redshift of $z \geq 2.5$ and a radio flux density of more than 50\,mJy, taken from the Roma BZCAT Multifrequency Catalogue of Blazars\footnote{\url{http://www.asdc.asi.it/bzcat/}}
and the SHAO list of high-$z$ radio-loud quasars\footnote{\url{http://202.127.29.4/CRATIV/en/high_z.html}}.
Sources with a high radio flux density are selected due to
a tight correlation between the radio and $\gamma$-ray flux \citep{Ghirlanda_2010,Ackermann_2011}.
The total sample consists of 169 BZCAT blazars, and seven SHAO sources. 

 \begin{figure}[pt]
   \centering
        \centering
    \includegraphics[width=1\hsize]{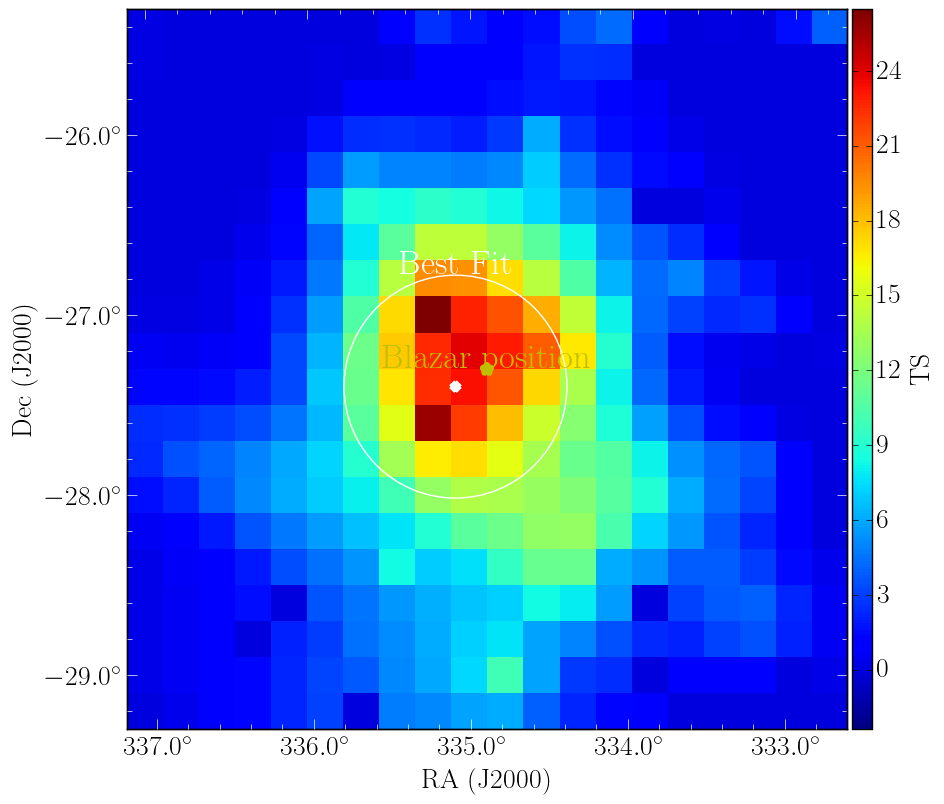}
    \caption{TS map of the blazar 5BZQ\,J2219$-$2719 at z = 3.63. The green star corresponds to the localization of the high-\textit{z} blazar using optical data. The white star and circle indicate the \textit{Fermi}/LAT best-fit position. An excess of  $\sim\,5\,\sigma$ is seen, consistent with the target blazar. }
  \label{TS map 1}
  \end{figure}

\subsection{\textit{Fermi}/LAT data analysis}\label{LAT_analysis}
\hspace{-13pt}
For the analysis of the \textit{Fermi}/LAT data we performed a dedicated likelihood and localization analysis of all blazars selected in Sec.\,\ref{Sample}.
Monthly binned \textit{Fermi}/LAT $\gamma$-ray light curves in the time range from 2008 August 4 to 2019 April 4  and energy range of 100\,MeV to 300\,GeV  were calculated, resulting in a total of 130 monthly intervals per blazar.
We used the \textit{Fermi} Science Tools (v11r5p3) together with the reprocessed Pass 8 data and the $\text{P8R3}\_\text{SOURCE}\_\text{V2}$ instrument response functions.
We performed an unbinned likelihood analysis in a region of interest (ROI) of $10^{\circ}$ around the source of interest for each of these time bins. 
A zenith angle cut of $90^{\circ}$ was used together with the LAT EVENT\_CLASS = 128 and the LAT EVENT\_TYPE = 3, while the \emph{gtmktime} cuts DATA\_QUAL==1 $\&\&$ LAT\_CONFIG==1 were chosen. For the modeling of the diffuse components\footnote{\url{https://fermi.gsfc.nasa.gov/ssc/data/access/lat/BackgroundModels.html}} we used the Galactic diffuse emission model gll\_iem\_v07.fits and the isotropic diffuse model iso\_P8R3\_SOURCE\_V2\_v01.txt.
The model used in the likelihood analysis contained all of the sources from the Fourth \textit{Fermi}/LAT Source catalog \citep[4FGL;][]{4FGL} within a radius of $20^{\circ}$ ($\text{ROI}+10^{\circ}$) from the source of interest.
In case the source of interest is not listed in the 4FGL catalog a new source is added, using the radio coordinates. If the studied high-\textit{z} blazar is already known to the 4FGL, this catalog source is used in the model.
Sources were fitted with either a power-law or a log-parabola fit following the 4FGL catalog.
New sources are always fitted with a power-law model.
In the inner 10 degree from the source of interest, the parameters PREFACTOR and INDEX were left free to vary for each monthly interval, while for sources further outside all parameters were fixed to the 4FGL values.
 The time binning of 30 days was chosen to study periods which would not have been monitored by the \textit{Fermi}/LAT Flare Advocate service \citep{FA_paper} while keeping the computing effort at a reasonable level.
As a measurement of the detection significance of each monthly interval, we calculated the test statistic following Wilks' theorem \citep{Wilks_1938,Mattox_1996}.
In cases where no significant detection was found, upper limits were calculated based on the method of \cite{Feldman_1998}.

 \begin{figure}[pt]
   \centering
        \centering
    \includegraphics[width=1\hsize]{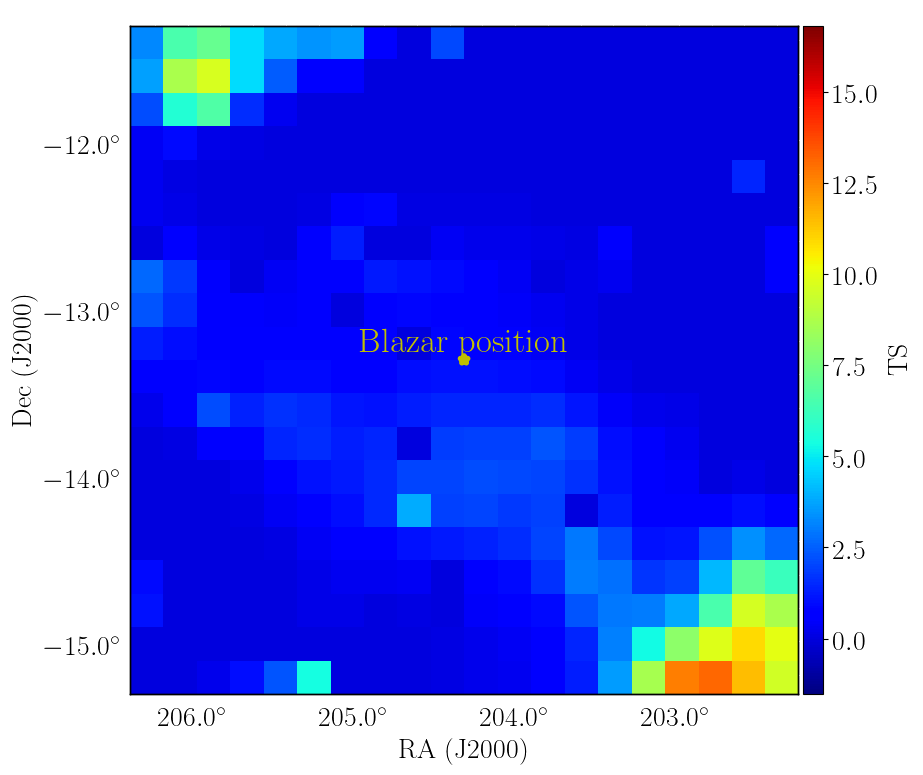}
    \caption{TS map of the blazar 5BZQ\,J1337$-$1319 at z = 3.48. The green star corresponds to the localization of the high-\textit{z} blazar using optical data. No excess at a position consistent with the target position is seen.}
  \label{TS map 2}
  \end{figure} 

Our localization analysis followed an iterative method using \emph{gtfindsrc}. For each monthly time range we started with the radio coordinates to find the best-fit position. Those coordinates were then used as input for the next localization step. This iteration was repeated until the final best-fit position varied by less than $0.05^{\circ}$.
In order to ensure that the observed emission could be associated to the previously undetected high-\textit{z} $\gamma$-ray blazar, we searched for additional possible counterparts within the localization region. Since the majority of monthly intervals was at a low significance level of TS\,$< 9$, we considered only intervals for which the best-fit position stayed stable. 
In this way, a contamination of the observed $\gamma$-ray flux from other sources was prevented.
TS maps were calculated for each monthly interval.
Figure\,\ref{TS map 1} shows such a map for the blazar 5BZQ\,J2219$-$2719 at z = 3.63. Monthly intervals for which no emission could be associated to the high-\textit{z} blazar best-fit position were not considered for further analysis. Fig.\,\ref{TS map 2} shows such an excluded interval for the blazar 5BZQ\,J1337$-$1319 at z = 3.48.
Such excluded intervals are caused by a misidentification from a bright nearby $\gamma$-ray emitter.

 \begin{figure}[pt]
  \centering
  \includegraphics[width=1\linewidth]{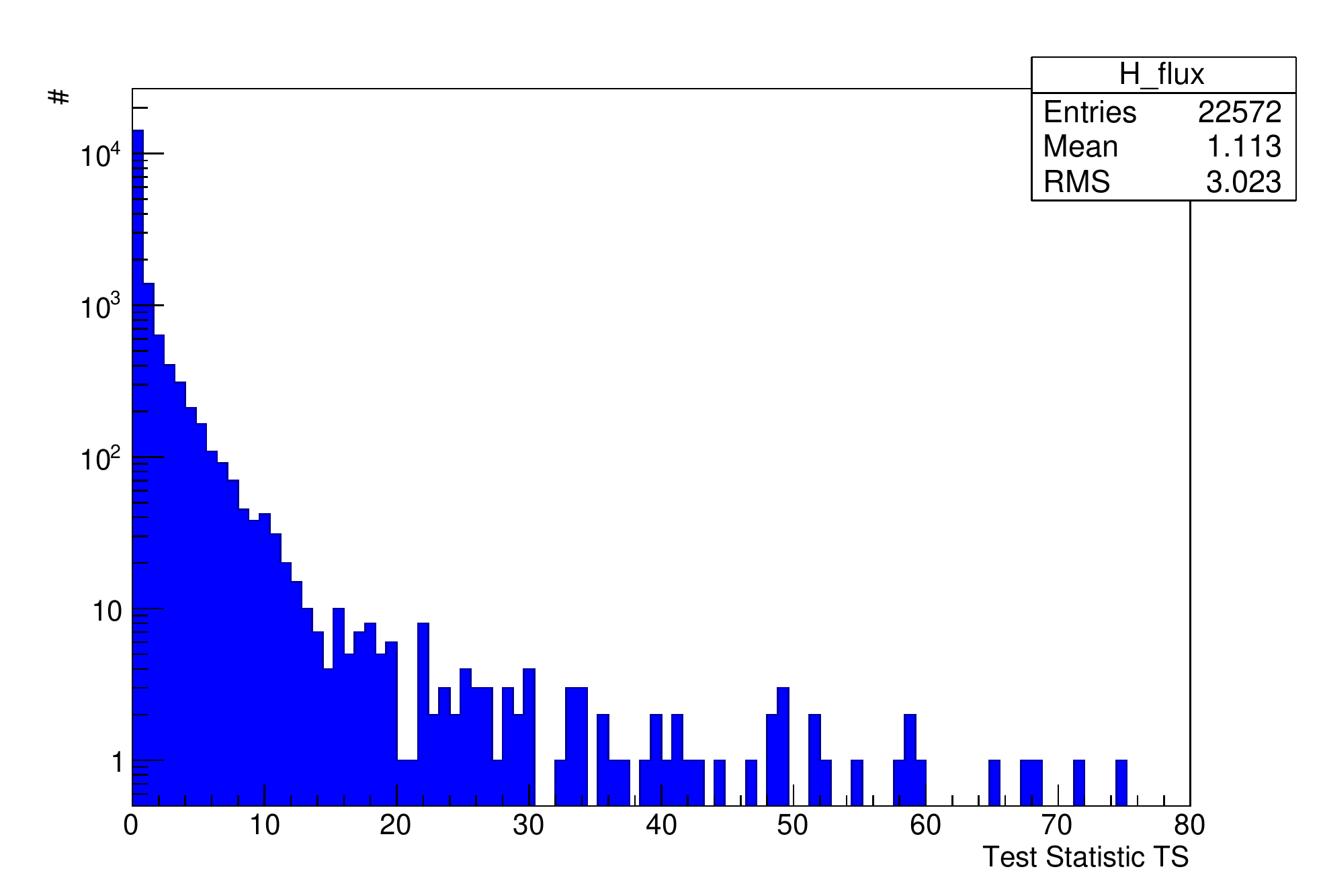} 
     \caption{Test statistic distribution of 176 blazar positions selected according to Sec.\,\ref{Sample}. 
     22 monthly intervals are identified with a significance of TS\,$\geq 25$, while at a  TS\,$\geq 9$ level
 249 intervals are detected. The TS axis is scaled from zero to 80.}
   \label{TS sources}
\end{figure}

\subsection{Background Determination}\label{Background determination}
\begin{figure}[pt]
   \centering
        \centering
    \includegraphics[width=1\hsize]{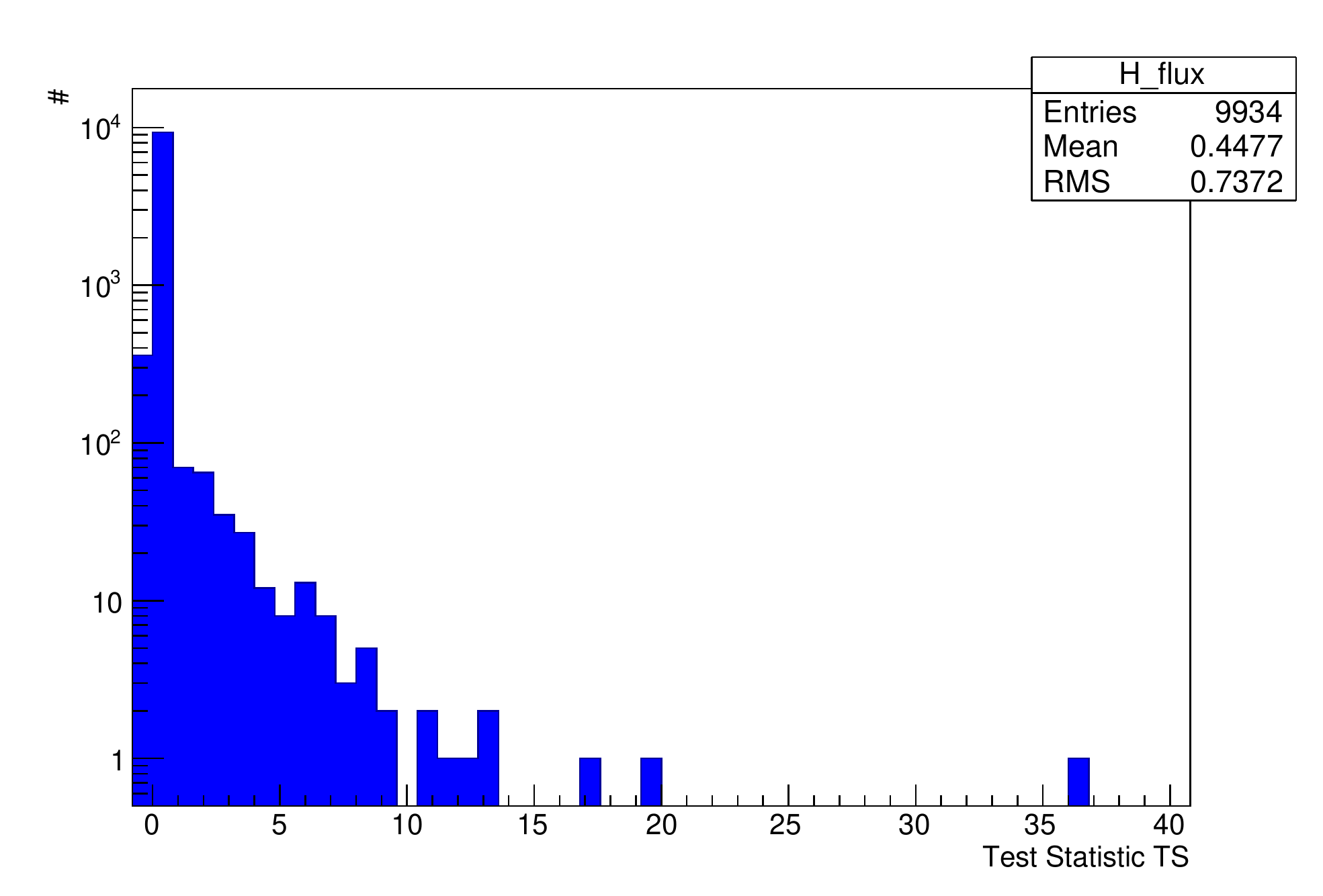}
    \caption{Test statistic distribution of a sample of 80 blank-sky positions. This histogram represents the background fluctuation for the performed \textit{Fermi}/LAT variability analysis. The TS axis is scaled from zero to 40.}
  \label{TS blank sky}
  \end{figure}

\hspace{-13pt}
At a test statistic level of TS\,$\geq\,25$ $\gamma$-ray detections are considered as trustworthy, because the chance for a false positive detection is about $5 \times 10^{-7}\,\%$.
Multiple (TS\,$<\,25$) detections from an unknown emitter, however, suggest the presence of a new $\gamma$-ray source, even if none of the monthly intervals are detected above the required (TS\,$>\,25$) threshold. In order to account for these sources, we measure the rate of false positive detections on monthly time scales over the entire $\gamma$-ray sky. This has not been done before. We perform a similar\footnote{The same energy, time range and binning are used.} \textit{Fermi}/LAT light-curve analysis on a sample of blank sky positions.
These positions have been randomly selected by mirroring the positions of known $\gamma$-ray blazars on the Galactic plane. In this way, the distribution of blank sky positions follows the observed blazar distribution in the sky.
To make sure the measured false positive rate is not affected by nearby $\gamma$-ray sources, we exclude all random positions which are less then $1.5^{\circ}$ apart from any known \textit{Fermi}/LAT source. Figure\,\ref{TS blank sky} shows the TS distribution of monthly intervals for the 80 blank-sky positions that were studied, while the TS distribution of the 176 high-$z$ blazars that were analyzed is shown in Fig.\,\ref{TS sources}.
From the total number of blank-sky intervals studied, only a fraction of $11/9934 \approx 1.1\times 10^{-3}$ monthly intervals showed a statistical fluctuation at a TS\,$\geq$\,9 level.
This corresponds to a Gaussian standard deviation of about $3.26\,\sigma$.
In order to derive a representative measurement of random fluctuations over the entire $\gamma$-ray sky, it is necessary to perform this measurement for an extremely large number of blank sky positions. To account for the (always) finite number of positions analyzed, we used a Poisson fit of the TS distribution in Fig.\,\ref{TS blank sky} to extrapolate the expected background rate to higher TS values and to account for TS values without measured monthly intervals. In this way, the effect of large TS values in Fig.\,\ref{TS blank sky}, which might be caused by a miss identification of a bright source in the vicinity of the blank-sky position is reduced. By comparing the (normalized) number of measured fluctuations $\mu_b$ to the (normalized) observed number of monthly intervals $N_s$ detected from our sample of radio and optically selected  high-$z$ blazars, we derive the Poisson probability $\Phi$ for a false positive detection of a detected monthly bin to be

\begin{align}
\Phi =  P\left(N_{s} \mid \mu_{b} \right).
\label{Phi Equation}
\end{align}

As a sanity check we solve Eqn.\,\ref{Phi Equation} for various test statistic levels.
The results can be seen in Table\,\ref{Tab. Phi}.
We find a sample size of 80 blank-sky positions sufficient to keep this estimate stable.
In order to determine whether the significance level of the signal measured
from a blazar which shows multiple monthly bins with a TS\,$< 25$ is sufficient to claim a detection, we consider the following two scenarios:
\begin{table}[pb]
    \caption{Poisson probability $\Phi$ of a single monthly light curve interval to be of purely statistical origin for various levels of test statistic.}
\centering
\begin{large}
    \begin{tabular}{c c}
    \label{Tab. Phi}
\textbf{TS value} & \textbf{$\Phi$ in $\sigma$}
\\\hline
10 & 3.26\\
15 & 3.60\\
25 & 4.88\\
55 & 7.26
 \\\hline
\end{tabular}
\end{large}
 \end{table}

  \begin{table*}[pt]
    \caption{The five high-z blazars that were identified, each having been detected
  in a single monthly interval with TS\,$\geq 25$. Three of our identified blazars are known as new $\gamma$-ray sources due to the recently published 4FGL catalog, while two previously unknown $\gamma$-ray 
     blazars have been identified. Four of the blazars found to have TS\,$\geq 25$ in at least one month were in the 3FGL and are therefore not listed in this table.
     The blazar 4FGL J0539.9$-$2839 has been included as a sanity check for the background estimate used.}
\centering
\begin{footnotesize}
    \begin{tabular}{c c c c c c c}
    \label{High-Z sources subsample}
\begin{tabular}{c} \textbf{Source} \\  \textbf{Name}     \end{tabular}    &    \begin{tabular}{c} \textbf{RA} \\  \textbf{J2000}     \end{tabular} &  \begin{tabular}{c} \textbf{DEC} \\  \textbf{J2000}     \end{tabular}  & \textbf{z}  & \begin{tabular}{c} \textbf{Detections} \\  \textbf{TS\,$\geq \text{9}$}     \end{tabular} &  \begin{tabular}{c} \textbf{Detections} \\  \textbf{TS\,$\geq \text{25}$}     \end{tabular} &  \begin{tabular}{c} \textbf{$\gamma$-ray} \\ \textbf{Counterpart} \end{tabular}\\\\\hline
5BZQ\,J2219$-$2719 &	334.90 &	-27.32 &	3.63 &	3 &	1 &	-/- \\
5BZQ\,J0422$-$3844 &	65.56 &	-38.75 &	3.11 &	7 &	1 &	-/- \\
5BZQ\,J0733+0456 &	113.49 &	4.94 &	3.01 &	19 &	7 &	4FGL J0733.8+0455 \\
5BZQ\,J1127+5650 &	171.92 &	56.84 &	2.89 &	14 &	3 &	4FGL J1127.4+5648 \\
5BZQ\,J0434$-$4355 &	68.51	 &	-43.93 &	2.65 &	9 &	2 &	4FGL J0440.3$-$4333 \\\hline
5BZQ\,J0539$-$2839&	84.98 &	-28.67 &3.10 &68 &	36 &	4FGL J0539.9$-$2839 \\
\hline
\end{tabular}
\end{footnotesize}
 \end{table*}

\subsubsection{Number of detections expected at  TS $\geq 9$} \label{How many P9}
\hspace{-13pt}
The probability that the entire observed $\gamma$-ray emission of individual blazars is purely below a detection threshold of TS $\geq 9$ is
\begin{align}
    P_{\ge N_s} = \sum\limits_{n = N_s}^{N} (P_9)^n * (1 - P_9)^{N - n} \begin{pmatrix} N \\ n \end{pmatrix}.
    \label{P9}
\end{align}
$P_9 = 1.1 \times 10^{-3}$ corresponds to the probability of a detection at a $3\,\sigma$ level, while $N$ indicates the total number of monthly intervals per light curve and $N_s$ sets the number of TS $\geq 9$ intervals. 
While the first two terms represent the number of light curve intervals observed with a \mbox{TS $\geq 9$} and vice versa, the last term $\begin{pmatrix} N \\ n \end{pmatrix}$ takes the number of possible combinations with identical outcome into account.  
In this calculation, the exact TS value of each light curve interval is not considered. 
In the following, we choose the blazar BZQ\,J1430+4204 (z = 4.72) listed in Table\,\ref{High-Z sources TS 9} as an example of the statistical method used. This blazar has been identified in five monthly light curve intervals up to TS $\sim 20$. Following Eqn.\,\ref{P9}, this results in a Poisson probability of $P_{\ge N_s} = 3.72\,\sigma$. This probability does not correspond to the detection significance of this blazar, but rather to the probability of how likely it is that this source in general is detected at a TS = 9 level.
Table\,\ref{Tab. Ns} lists the Poisson probability $P_{\ge N_s}$ for different numbers of TS $\geq 9$ intervals.
   
\subsubsection{Number of detections expected at each TS value}
\hspace{-13pt}
To take the exact TS value of each light-curve interval into account, Eqn.\,\ref{P9} is modified to

\begin{align}
    P_{tot} =  \sum\limits_{i = 1}^{N} \Phi_{i} * (1 - \Phi_{i})^{N-1} * 100
    \label{P9_TS_value}
\end{align}
Here, $\Phi_i$ corresponds to the Poisson probability calculated following Eqn.\,\ref{Phi Equation}. 
The combinatoric factor is reduced to 100, as each   
combination is unique and can occur at any of the bins of the blazar TS distribution. 
Applying Eqn.\,\ref{P9_TS_value} to BZQ\,J1430+4204 leads to a total probability of $P_{tot} = 6.57\,\sigma$. In this calculation the exact TS value of each light curve interval has been 
considered. 
\begin{table}[pb]
    \caption{Poisson probability $P_{\ge N_s}$ following Eq.\,\ref{P9} for different numbers of TS $\geq 9$ intervals.}
\centering
\begin{large}
    \begin{tabular}{c c}
    \label{Tab. Ns}
$\text{\textbf{N}}_\text{\textbf{s}}$ & \textbf{$P_{\ge N_s}$ in $\sigma$}
\\\hline
1 & 3.26\\
2 & 3.38\\
3 & 3.49\\
4 & 3.61 \\
\end{tabular}
\end{large}
 \end{table}

By fixing $\Phi_i = P_9$ and allowing for multiple combinations, our result becomes equal to the probability derived in Sec.\,\ref{How many P9}. Nevertheless, $P_{tot}$ represents the total probability of observing $\gamma$-ray emission from the studied blazar, considering the exact TS level of the measured $\gamma$-ray emission.

\subsection{\textit{Swift}/XRT data analysis}
\hspace{-13pt}
In our sample of significantly detected new $\gamma$-ray blazars listed in Table\,\ref{High-Z sources TS 9}, 5BZQ\,J1429+5406 is the only source for which \textit{Swift}/XRT data are available.~These data were reduced with standard methods, using the software package HEASOFT v. 6.26.1. The data were reduced, calibrated and cleaned by means of the XRT-PIPELINE script using standard filtering criteria. We selected 0-12 grades for observations in photon counting (pc) mode. Spectral fitting was performed with ISIS 1.6.2 \citep{Houck2000}.
The data for 5BZQ\,J1429+5406  were extracted using XSELECT with a source radius of
54.218$^{\prime\prime}$ centered on $\alpha=$217.60, $\delta=$42.08, in order to cover $\gtrsim 90\%$ of the PSF at all energies.~The background region was extracted from an annulus centered on the same coordinates, with radii of 82.506$^{\prime\prime}$ and 235.731$^{\prime\prime}$, while ensuring that no X-ray source was present in the background region.
5BZQ\,J1429+5406 shows a count rate of $0.04$\,cts/s, which is not bright enough to raise any concerns about pile-up.
Unfortunately, these data were not simultaneous with the observed $\gamma$-ray emission.

\section{Individual high-\textit{z} blazars}\label{Individual Intervalls}

\subsection{TS $\geq 25$ detections}
~\hspace{-19pt}
Out of our sample of 176  high-$z$ blazars, five were detected in at least one monthly interval with a test statistic of TS\,$\geq 25$.
Table\,\ref{High-Z sources subsample} lists the relevant properties of these blazars. Three blazars showed evidence for stronger emission in the first eight years of the \textit{Fermi} mission and were, therefore, included in the 4FGL catalog.
Two new high-$z$ blazars were identified as $\gamma$-ray blazars by the analysis performed here.
They each showed evidence for heightened activity (TS\,$> 25$) during one of the monthly
time bins in the period after the eight-year time range considered for the 4FGL catalog.
As can be seen in Table\,\ref{High-Z sources subsample}, the source 5BZQ\,J2219$-$2719, at a redshift of z = 3.63, is identified as the most distant TS\,$>25$ blazar detected in our work. Besides their emission at a TS level of more then 25, these blazars showed multiple monthly intervals at a TS\,$\geq\,9$ level. 
As a proof of principle of the performed background estimate, we included the  known blazar PKS\,0537$-$286 (z = 3.10) in our sample. This source has been detected multiple times at different significance levels on monthly time scales and even showed
daily flaring activity \citep{Cheung_2017}. Following Eqn.\,\ref{P9_TS_value} for all monthly intervals with a TS\,$\geq\,9$, we show that the Poisson probability that these detections are of purely random origin can be excluded at least with $P_{tot} \sim 8\,\sigma$.
This further proves that blazars at these large distances can be prominent $\gamma$-ray emitters.

\subsection{TS\,$\geq$\,9 detections}
~\hspace{-17pt}
At a TS\,$\geq$\,9 level, a total of 249 monthly intervals have been identified from 116 high-$z$ blazars.
This suggests that the majority of blazars studied in this work show at least indications of $\gamma$-ray flaring activity.
 By comparing the individual TS distribution of the 116 TS\,$\geq$\,9 blazars to the measured TS distribution of random fluctuations seen in Fig.\,\ref{TS blank sky} we find that 27 previously $\gamma$-ray undetected blazars show multiple monthly detections above the expected background.
Table\,\ref{High-Z sources TS 9} lists the relevant properties of these blazars.
 While four of these blazars are in the 4FGL catalog due to increased
levels of $\gamma$-ray emission after the time range considered in the 3FGL, 23 previously unknown high-$z$ blazar candidates have been identified, each  with a total probability, $P_{tot}$, of at least $3\,\sigma$. 
 Out of these 23 potential new $\gamma$-ray blazars, six are detected with a $P_{tot} > 5\,\sigma$. 
 In this sample of potential $\gamma$-ray blazars (TS\,$\geq 9$ in more than
one monthly interval)  at high redshift, 16 previously $\gamma$-ray undetected blazars are located at a redshift of z\,$\geq$\,3 with 6 of
these blazars being even more distant at a redshift of z\,$\geq$\,4.  
 The most distant blazars in our sample 5BZQ\,J0906+6930 (z =  5.47) and 5BZQ\,J1026+2542 (z = 5.29) had a signal at the TS\,$\sim 10$ level during just one monthly interval each
in the time period analyzed here. This is consistent with background fluctuations and therefore these sources are only considered as promising $\gamma$-ray candidates. 

\section{Discussion} \label{Discussion}
\hspace{-13pt}
In this section, we compare our sample of new $\gamma$-ray blazars observed on monthly time scales to the first catalog of \textit{Fermi}/LAT transients (1FLT), discuss limitations of the method used and present the $\gamma$-ray properties of these new high-energy emitters.

\subsection{Comparison to 1FLT catalog}
\hspace{-13pt}
The 1FLT represents the first catalog of $\gamma$-ray transient sources detected on monthly time scales by \textit{Fermi}/LAT. 
While general source catalogs like the 4FGL integrate over the entire mission time of the \textit{Fermi} satellite \citep{4FGL}, the 1FLT searches for TS$> 25$ detections on shorter time scales.
Therefore, transient sources listed in the 1FLT are in general unknown to the long-term integrated catalogs.\\
Our results listed in Table\,\ref{High-Z sources subsample} are fully consistent with the findings of the 1FLT catalog.~In fact, only the new $\gamma$-ray detected blazar 5BZQ\,J2219$-$2719 is listed in the 1FLT, as this catalog does not include sources listed in the 4FGL and only covers a time range up to August 2018.

\subsection{Spectral properties}\label{Spectral properties}
\hspace{-13pt}
Little is known about the radiative mechanisms of $\gamma$-ray loud blazars at high redshifts. In contrast to blazars in the local Universe, the SEDs of their distant relatives seem to be dominated by their high-energy peak in the blazar SED, which rises above the synchrotron peak by approximately one order of magnitude \citep{Celotti_2008}.
To understand the physical processes, which power this luminous subclass of distant AGN, multiwavelength studies of their spectral properties are crucial.~Monthly $\gamma$-ray detections, as can be seen in Fig.\,\ref{PKS0438LC}, have so far only been discovered months or even years after the detections, making simultaneous observations in other wavebands
impossible.~Building broadband SEDs for most high-\textit{z} blazars is therefore extremely challenging as the number of sources detected at different wavelengths is very limited. In fact, out of all new significantly detected $\gamma$-ray blazars\footnote{This includes the two new sources in Table\,\ref{High-Z sources subsample} as well as the six blazars in Table\,\ref{High-Z sources TS 9} which show a  $P_{tot} > 5\,\sigma$.} listed in Table\,\ref{High-Z sources subsample} and Table\,\ref{High-Z sources TS 9}, only a single blazar was observed at both X-ray and  $\gamma$-ray energies. Figure\,\ref{SED} displays the high-energy hump of the blazar 5BZQ\,J1430+4204, using \textit{Swift}/XRT and \textit{Fermi}/LAT data.\\
The \textit{Fermi}/LAT spectrum shows a soft power-law spectral index of $2.7$, which is also common for blazars in the local universe during flaring states. The entire $\gamma$-ray emission is centered in the first spectral bin with TS\,$\sim 12$. Such behavior is expected for $\gamma$-ray detections of high-\textit{z} blazars due to EBL scattering. \\
\textit{Swift}/XRT data from 5BZQ\,J1430+4204 were fit using Cash statistics with a 
S/N binning of 1 and an energy range of 0.3--10\,keV.
Fitting an absorbed power law yields a best-fit statistics of 
$\chi^{2}_{\mathrm{reduced}}=0.88$ with 232 degrees of freedom.
The best-fit power-law index is $\Gamma=1.18^{+0.12}_{-0.10}$, while the 
absorption has been fixed to the Galactic value of 9.34$\times 
10^{19}$\,cm$^{-2}$. The unabsorbed 2--10\,keV flux is 
$1.73^{+0.25}_{-0.23}\times10^{-12}$\,erg\,s$^{-1}$\,cm$^{-2}$. At 
$z=4.72$, this corresponds to a luminosity of 
$3.9\pm0.6\times10^{47}$\,erg\,s$^{-1}$.

\subsection{Limitations of the method used}

\subsubsection{Sample selection}
In this work, we presented a targeted search for $\gamma$-ray emission at high redshifts, based on a sample of 176 radio and optically selected  blazars. Furthermore, only blazars with a certain radio flux density were selected\footnote{Please refer to Sec.\,\ref{Sample} for further details.}. This introduces a bias in the nature of detectable high-\textit{z} blazars and limits the potential of possible $\gamma$-ray detections. A more general search for $\gamma$-ray emitting high-\textit{z} blazars, including a trigger for quasi-simultaneous multiwavelength data will be the scope of our future work.
\hspace{-13pt}


\subsubsection{Significance calculation}
\hspace{-13pt}
In order to calculate the significance of a new blazar detection based on multiple TS\,$< 25$ intervals, we systematically measured the amount of random fluctuations in the entire $\gamma$-ray sky on monthly time scales. To account for the (always) finite number of blank-sky positions studied, we normalize the TS distributions of our high-\textit{z} blazar sample seen in Fig.\,\ref{TS sources} as well as the TS distribution of the blank-sky positions studied, as seen in Fig.\,\ref{TS blank sky}. Furthermore, we fit these distributions with a Poisson function\footnote{Please refer to Sec.\,\ref{Background determination} for further details.} to account for TS values that have not been measured.\\
To calculate the uncertainty of the total detection probability $\Delta\,P_{tot}$, we vary the parameters of the Poisson fit. We also tested different functions (Gaussian and exponential) to quantify how well they described the measured TS distributions. We find that the total detection probability $P_{tot}$ differs by about $2\,\%$ to $10\,\%$ for different fit functions or parameters. Thus, we estimate the systematic uncertainty of the presented significance calculation to be on the order of $\sim\,10\,\%$.

\subsubsection{Counterpart association}
\hspace{-13pt}
Within the  localization uncertainty of the high-\textit{z} blazars studied (see e.g. Fig.\,\ref{TS map 1}), additional $\gamma$-ray counterparts could also be present. Since we  only consider multiple light-curve intervals for those objects whose best-fit positions stay
stable at the same coordinates, a contamination from unrelated $\gamma$-ray emitters within the localization uncertainty seems to be unlikely.
However, even if the positional variance is low, a misidentification of other $\gamma$-ray emitting sources consistent with the best-fit position can not be fully ruled out. Therefore, we consider the identified high-\textit{z} blazars as the most plausible counterparts to the observed $\gamma$-ray emission.

\section{Conclusion}\label{Conclusion}
\hspace{-15pt}
We have studied the long-term MeV -- GeV variability behavior of a sample of 176 radio and optically detected high-\textit{z} blazars using \textit{Fermi}/LAT. A dedicated localization analysis has been performed to ensure an association between the observed $\gamma$-ray emission and the targeted high-\textit{z} blazars. Indications of $\gamma$-ray emission have been detected from more than half of the blazar sample studied.~The two previously unknown  $\gamma$-ray blazars 5BZQ\,J2219$-$2719 (z = 3.63) and 5BZQ\,J0422$-$3844 (z = 3.11) have each been identified in a single monthly interval at a significance level of TS\,$\geq\,25$. In addition, we have confirmed, at the TS\,$> 25$ level, $\gamma$-ray
emission from three 4FGL blazars on monthly timescales. These only
showed enhanced $\gamma$-ray activity in recent years and were, therefore, not listed in the 3FGL catalog. 
In order to account for the huge detection potential of multiple TS$< 25$ light curve intervals, we have systematically studied the expected number of all-sky random fluctuations on monthly time scales. To calculate the total detection probability $P_{tot}$ (see Eqn.\,\ref{P9_TS_value}) of multiple TS $< 25$ detections, we compared the number of measured light curve intervals of the blazars considered to the number of expected monthly fluctuations at each TS value. Table\,\ref{High-Z sources TS 9} lists the results for 27 blazars which have been identified in multiple low-significance detections.
This includes 23 blazars previously undetected at $\gamma$-ray energies.
Four of these blazars were not in the 3FGL catalog but were included in 4FGL, the most recent
\textit{Fermi}/LAT source catalog.
This demonstrates our method to be robust, as multiple low-significance detections provide a clear indication for a new $\gamma$-ray source.\\
Six blazars listed in Table\,\ref{High-Z sources TS 9} have been identified with a $P_{tot} \geq 5\,\sigma$, and are marked in bold. These include the most distant blazars studied in our sample. The two blazars 5BZQ\,J1430+4204 (z = 4.72) and 5BZU\,J0525$-$3343 (z = 4.41) lie at redshifts more distant than previously reported $\gamma$-ray blazars, making them the most distant blazars ever detected by \textit{Fermi}/LAT. In contrast to the blazars reported by \cite{Ackermann_2017}, which were detected in an integrated analysis over 92 months of LAT data, the long-term integrated flux of our sources is too faint for a significant source
detection over long time scales.\\
Our results are consistent with the findings of \cite{Liao_2018}, who independently confirmed $\gamma$-ray emission observed by \textit{Fermi}/LAT from the direction of the blazar 5BZQ\,J1430+4204. While \cite{Liao_2018} claim a flaring period of about 10 months in
the time range of July 2012 to May 2013, our variability analysis
indicates that it was only during three of the monthly intervals in
this time period that 5BZQ\,J1430+4204 showed evidence for $\gamma$-ray emission at TS\,$\sim 12$.
Furthermore, we also find indications of enhanced $\gamma$-ray activity in monthly bins at a TS\,$\sim 15$ level covering a time range from December 2014 to January 2015. 

 \begin{figure}[pt]
  \centering
  \includegraphics[width=1\linewidth]{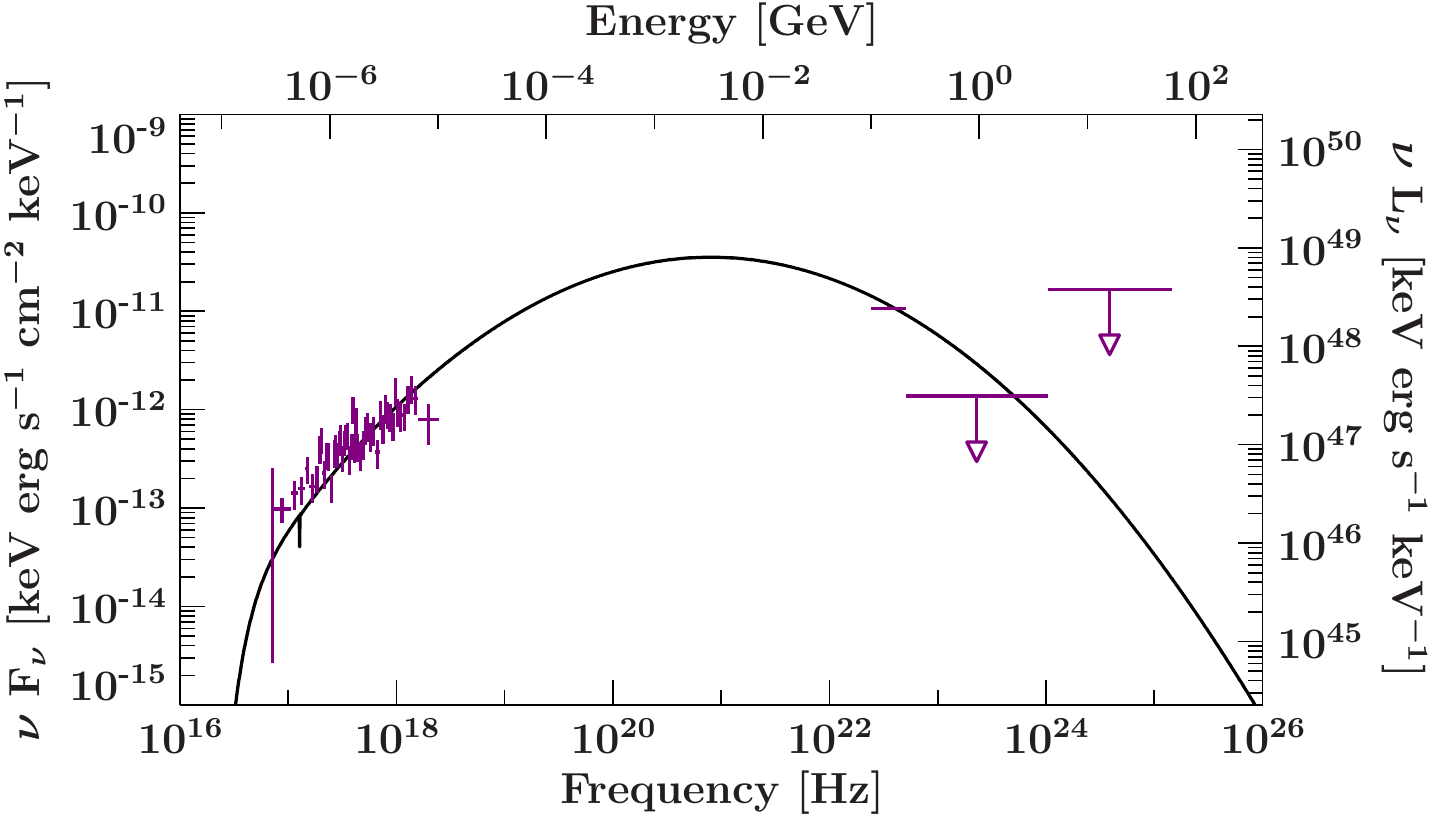} 
     \caption{Multiwavelength SED of 5BZQ\,J1430+4204 (z = 4.72), using \textit{Swift}/XRT and \textit{Fermi}/LAT data. The blazar high-energy hump is modeled with a log parabola. The entire emission observed by \textit{Fermi}/LAT is centered on the $\sim$100\,MeV bin with a test statistic of TS\,$\sim 12$.}
   \label{SED}
\end{figure}

This differences in both analyses are likely due to the use of different sky models and instrument response functions. While \cite{Liao_2018} use the preliminary \textit{Fermi}/LAT 8 year Source List (FL8Y\footnote{\url{https://fermi.gsfc.nasa.gov/ssc/data/access/lat/fl8y/}}), our work is based on the most recent general source catalog (4FGL).
In an independent analysis, \cite{Li_2020} confirmed transient $\gamma$-ray activity from the two blazars  5BZQ\,J2219$-$2719 and 5BZQ\,J2321$-$0827, respectively. These sources account for two of the most significant detected new blazars in our sample. By studying flaring periods of several months, \cite{Li_2020} was able to identify both blazars with a TS\,$>25$.\\
In addition to the previously discussed high-\textit{z} blazar identifications, indications of high-energy emission at a $P_{tot} > 4\,\sigma$ level have been identified from 36 blazars, including 16 blazars previously undetected by \textit{Fermi}/LAT.
Two blazars, 5BZQ\,J0906+6930 (z = 5.47) and 5BZQ\,J1026+2542 (z = 5.29), each showed evidence for emission during a single monthly time interval with a TS of $\sim 10$. Since our analysis has shown that single detections at this low-significance level are fully consistent
with random fluctuations, we do not consider these two blazars as $\gamma$-ray emitters.\\ 
In Sec.\,\ref{Spectral properties},  we studied the spectral properties of our most distant significantly detected  blazar 5BZQ\,J1430+4204 (z = 4.72). Figure\,\ref{SED} displays the high-energy peak of the blazar SED, using \textit{Swift}/XRT and \textit{Fermi}/LAT data.
The X-ray through $\gamma$-ray SED is very poorly constrained, but does not show obvious signs of unusual high-energy spectral components different from the high-energy SEDs in lower-redshift blazars \citep{Fan_2016}.\\
Due to the sparse multiwavelength coverage of most of the blazars studied, (quasi)-simultaneous SEDs could not be constructed. 
Due the the large distance of these sources, however, the $\gamma$-ray emission peaks towards longer wavelengths, causing a more pronounced increase in the emission in the keV -- MeV regime.
Future missions like the All-sky Medium Energy Gamma-ray Observatory (AMEGO) will provide a substantial contribution to the study of blazars in the early Universe \citep{Paliya_2019b}.  
Quasi-simultaneous multiwavelength data of high-\textit{z} blazars would provide crucial insights into the spectral properties and emission behaviors of blazars which existed so early in the Universe. A detailed study of (quasi)-simultaneous MWL observations of flaring high-\textit{z} blazars will be the focus of our future work.

\begin{acknowledgements}
\hspace{-14pt}
The work of M. Kreter and M. B\"ottcher is supported by the South African 
Research Chairs Initiative (grant no. 64789) of the Department of Science and 
Innovation and the National Research Foundation\footnote{Any opinion, finding 
and conclusion or recommendation expressed in this material is that of the authors 
and the NRF does not accept any liability in this regard.} of South Africa.
A. Gokus was partially funded by the Bundesministerium
für Wirtschaft und Technologie under Deutsches Zentrum für Luft- und Raumfahrt 
(grant 50OR1607O) and by the German Science Foundation (DFG, grant KR 3338/4-1).
F. Krau{\ss} was supported as an Eberly Research Fellow by the Eberly College of Science at the Pennsylvania State University.
 We thank J.E.~Davis for the development of the \texttt{slxfig}
  module that has been used to prepare the figures in this work.
  This research has made use of a collection of
  ISIS scripts provided by the Dr. Karl Remeis-Observatory, Bamberg,
  Germany at \url{http://www.sternwarte.uni-erlangen.de/isis/}.
The \textit{Fermi} LAT Collaboration acknowledges generous ongoing support
from a number of agencies and institutes that have supported both the
development and the operation of the LAT as well as scientific data analysis.
These include the National Aeronautics and Space Administration and the
Department of Energy in the United States, the Commissariat \`a l'Energie Atomique
and the Centre National de la Recherche Scientifique / Institut National de Physique
Nucl\'eaire et de Physique des Particules in France, the Agenzia Spaziale Italiana
and the Istituto Nazionale di Fisica Nucleare in Italy, the Ministry of Education,
Culture, Sports, Science and Technology (MEXT), High Energy Accelerator Research
Organization (KEK) and Japan Aerospace Exploration Agency (JAXA) in Japan, and
the K.~A.~Wallenberg Foundation, the Swedish Research Council and the
Swedish National Space Board in Sweden.
Additional support for science analysis during the operations phase is gratefully
acknowledged from the Istituto Nazionale di Astrofisica in Italy and the Centre
National d'\'Etudes Spatiales in France. This work performed in part under DOE
Contract DE-AC02-76SF00515.
  We acknowledge the use of public data from the Swift data archive. 
  We thank Vaidehi Paliya for providing the broadband SED data of DA 193 shown in Fig.\,\ref{SED_plot}.
\end{acknowledgements}
\bibliography{References.bib}

\begin{thebibliography}{}
\expandafter\ifx\csname natexlab\endcsname\relax\def\natexlab#1{#1}\fi

\bibitem[{{Abdollahi} {et~al.}(2020){Abdollahi}, {Acero}, {Ackermann},
  {Ajello}, {Atwood}, {Axelsson}, {Baldini}, {Ballet}, {Barbiellini},
  {Bastieri}, {Becerra Gonzalez}, {Bellazzini}, {Berretta}, {Bissaldi}, {Bland
  ford}, {Bloom}, {Bonino}, {Bottacini}, {Brandt}, {Bregeon}, {Bruel},
  {Buehler}, {Burnett}, {Buson}, {Cameron}, {Caputo}, {Caraveo}, {Casandjian},
  {Castro}, {Cavazzuti}, {Charles}, {Chaty}, {Chen}, {Cheung}, {Chiaro},
  {Ciprini}, {Cohen-Tanugi}, {Cominsky}, {Coronado-Bl{\'a}zquez}, {Costantin},
  {Cuoco}, {Cutini}, {D'Ammando}, {DeKlotz}, {Torre Luque}, {de Palma},
  {Desai}, {Digel}, {Lalla}, {Mauro}, {Venere}, {Dom{\'\i}nguez}, {Dumora},
  {Dirirsa}, {Fegan}, {Ferrara}, {Franckowiak}, {Fukazawa}, {Funk}, {Fusco},
  {Gargano}, {Gasparrini}, {Giglietto}, {Giommi}, {Giordano}, {Giroletti},
  {Glanzman}, {Green}, {Grenier}, {Griffin}, {Grondin}, {Grove}, {Guiriec},
  {Harding}, {Hayashi}, {Hays}, {Hewitt}, {Horan}, {J{\'o}hannesson},
  {Johnson}, {Kamae}, {Kerr}, {Kocevski}, {Kovac'evic'}, {Kuss}, {Landriu},
  {Larsson}, {Latronico}, {Lemoine-Goumard}, {Li}, {Liodakis}, {Longo},
  {Loparco}, {Lott}, {Lovellette}, {Lubrano}, {Madejski}, {Maldera},
  {Malyshev}, {Manfreda}, {Marchesini}, {Marcotulli}, {Mart{\'\i}-Devesa},
  {Martin}, {Massaro}, {Mazziotta}, {McEnery}, {Mereu}, {Meyer}, {Michelson},
  {Mirabal}, {Mizuno}, {Monzani}, {Morselli}, {Moskalenko}, {Negro}, {Nuss},
  {Ojha}, {Omodei}, {Orienti}, {Orlando}, {Ormes}, {Palatiello}, {Paliya},
  {Paneque}, {Pei}, {Pe{\~n}a-Herazo}, {Perkins}, {Persic}, {Pesce-Rollins},
  {Petrosian}, {Petrov}, {Piron}, {Poon}, {Porter}, {Principe}, {Rain{\`o}},
  {Rando}, {Razzano}, {Razzaque}, {Reimer}, {Reimer}, {Remy}, {Reposeur},
  {Romani}, {Parkinson}, {Schinzel}, {Serini}, {Sgr{\`o}}, {Siskind}, {Smith},
  {Spandre}, {Spinelli}, {Strong}, {Suson}, {Tajima}, {Takahashi}, {Tak},
  {Thayer}, {Thompson}, {Tibaldo}, {Torres}, {Torresi}, {Valverde}, {Klaveren},
  {Zyl}, {Wood}, {Yassine}, \& {Zaharijas}}]{4FGL}
{Abdollahi}, S., {Acero}, F., {Ackermann}, M., {et~al.} 2020, \apjs, 247, 33

\bibitem[{{Abolfathi} {et~al.}(2018){Abolfathi}, {Aguado}, {Aguilar}, {Allende
  Prieto}, {Almeida}, {Ananna}, {Anders}, {Anderson}, {Andrews}, {Anguiano},
  {Arag{\'o}n-Salamanca}, {Argudo-Fern{\'a}ndez}, {Armengaud}, {Ata},
  {Aubourg}, {Avila-Reese}, {Badenes}, {Bailey}, {Balland}, {Barger},
  {Barrera-Ballesteros}, {Bartosz}, {Bastien}, {Bates}, {Baumgarten},
  {Bautista}, {Beaton}, {Beers}, {Belfiore}, {Bender}, {Bernardi}, {Bershady},
  {Beutler}, {Bird}, {Bizyaev}, {Blanc}, {Blanton}, {Blomqvist}, {Bolton},
  {Boquien}, {Borissova}, {Bovy}, {Bradna Diaz}, {Brandt}, {Brinkmann},
  {Brownstein}, {Bundy}, {Burgasser}, {Burtin}, {Busca}, {Ca{\~n}as},
  {Cano-D{\'\i}az}, {Cappellari}, {Carrera}, {Casey}, {Cervantes Sodi}, {Chen},
  {Cherinka}, {Chiappini}, {Choi}, {Chojnowski}, {Chuang}, {Chung}, {Clerc},
  {Cohen}, {Comerford}, {Comparat}, {Correa do Nascimento}, {da Costa},
  {Cousinou}, {Covey}, {Crane}, {Cruz-Gonzalez}, {Cunha}, {da Silva Ilha},
  {Damke}, {Darling}, {Davidson}, {Dawson}, {de Icaza Lizaola}, {de la
  Macorra}, {de la Torre}, {De Lee}, {de Sainte Agathe}, {Deconto Machado},
  {Dell'Agli}, {Delubac}, {Diamond-Stanic}, {Donor}, {Downes}, {Drory}, {du Mas
  des Bourboux}, {Duckworth}, {Dwelly}, {Dyer}, {Ebelke}, {Davis Eigenbrot},
  {Eisenstein}, {Elsworth}, {Emsellem}, {Eracleous}, {Erfanianfar},
  {Escoffier}, {Fan}, {Fern{\'a}ndez Alvar}, {Fernandez-Trincado}, {Fernand o
  Cirolini}, {Feuillet}, {Finoguenov}, {Fleming}, {Font-Ribera}, {Freischlad},
  {Frinchaboy}, {Fu}, {G{\'o}mez Maqueo Chew}, {Galbany}, {Garc{\'\i}a
  P{\'e}rez}, {Garcia-Dias}, {Garc{\'\i}a-Hern{\'a}ndez}, {Garma Oehmichen},
  {Gaulme}, {Gelfand }, {Gil-Mar{\'\i}n}, {Gillespie}, {Goddard}, {Gonz{\'a}lez
  Hern{\'a}ndez}, {Gonzalez-Perez}, {Grabowski}, {Green}, {Grier}, {Gueguen},
  {Guo}, {Guy}, {Hagen}, {Hall}, {Harding}, {Hasselquist}, {Hawley}, {Hayes},
  {Hearty}, {Hekker}, {Hernand ez}, {Hernandez Toledo}, {Hogg},
  {Holley-Bockelmann}, {Holtzman}, {Hou}, {Hsieh}, {Hunt}, {Hutchinson},
  {Hwang}, {Jimenez Angel}, {Johnson}, {Jones}, {J{\"o}nsson}, {Jullo}, {Khan},
  {Kinemuchi}, {Kirkby}, {Kirkpatrick}, {Kitaura}, {Knapp}, {Kneib},
  {Kollmeier}, {Lacerna}, {Lane}, {Lang}, {Law}, {Le Goff}, {Lee}, {Li}, {Li},
  {Lian}, {Liang}, {Lima}, {Lin}, {Long}, {Lucatello}, {Lundgren}, {Mackereth},
  {MacLeod}, {Mahadevan}, {Maia}, {Majewski}, {Manchado}, {Maraston},
  {Mariappan}, {Marques-Chaves}, {Masseron}, {Masters}, {McDermid}, {McGreer},
  {Melendez}, {Meneses-Goytia}, {Merloni}, {Merrifield}, {Meszaros}, {Meza},
  {Minchev}, {Minniti}, {Mueller}, {Muller-Sanchez}, {Muna}, {Mu{\~n}oz},
  {Myers}, {Nair}, {Nand ra}, {Ness}, {Newman}, {Nichol}, {Nidever},
  {Nitschelm}, {Noterdaeme}, {O'Connell}, {Oelkers}, {Oravetz}, {Oravetz},
  {Ort{\'\i}z}, {Osorio}, {Pace}, {Padilla}, {Palanque-Delabrouille},
  {Palicio}, {Pan}, {Pan}, {Parikh}, {P{\^a}ris}, {Park}, {Peirani},
  {Pellejero-Ibanez}, {Penny}, {Percival}, {Perez-Fournon}, {Petitjean},
  {Pieri}, {Pinsonneault}, {Pisani}, {Prada}, {Prakash}, {Queiroz}, {Raddick},
  {Raichoor}, {Barboza Rembold}, {Richstein}, {Riffel}, {Riffel}, {Rix},
  {Robin}, {Rodr{\'\i}guez Torres}, {Rom{\'a}n-Z{\'u}{\~n}iga}, {Ross},
  {Rossi}, {Ruan}, {Ruggeri}, {Ruiz}, {Salvato}, {S{\'a}nchez}, {S{\'a}nchez},
  {Sanchez Almeida}, {S{\'a}nchez-Gallego}, {Santana Rojas}, {Santiago},
  {Schiavon}, {Schimoia}, {Schlafly}, {Schlegel}, {Schneider}, {Schuster},
  {Schwope}, {Seo}, {Serenelli}, {Shen}, {Shen}, {Shetrone}, {Shull}, {Silva
  Aguirre}, {Simon}, {Skrutskie}, {Slosar}, {Smethurst}, {Smith}, {Sobeck},
  {Somers}, {Souter}, {Souto}, {Spindler}, {Stark}, {Stassun}, {Steinmetz},
  {Stello}, {Storchi-Bergmann}, {Streblyanska}, {Stringfellow}, {Su{\'a}rez},
  {Sun}, {Szigeti}, {Taghizadeh-Popp}, {Talbot}, {Tang}, {Tao}, {Tayar},
  {Tembe}, {Teske}, {Thakar}, {Thomas}, {Tissera}, {Tojeiro}, {Tremonti},
  {Troup}, {Urry}, {Valenzuela}, {van den Bosch}, {Vargas-Gonz{\'a}lez},
  {Vargas-Maga{\~n}a}, {Vazquez}, {Villanova}, {Vogt}, {Wake}, {Wang},
  {Weaver}, {Weijmans}, {Weinberg}, {Westfall}, {Whelan}, {Wilcots}, {Wild},
  {Williams}, {Wilson}, {Wood-Vasey}, {Wylezalek}, {Xiao}, {Yan}, {Yang},
  {Ybarra}, {Y{\`e}che}, {Zakamska}, {Zamora}, {Zarrouk}, {Zasowski}, {Zhang},
  {Zhao}, {Zhao}, {Zheng}, {Zheng}, {Zhou}, {Zhu}, {Zinn}, \&
  {Zou}}]{Abolfathi_2018}
{Abolfathi}, B., {Aguado}, D.~S., {Aguilar}, G., {et~al.} 2018, \apjs, 235, 42

\bibitem[{{Acero} {et~al.}(2015){Acero}, {Ackermann}, {Ajello}, {Albert},
  {Atwood}, {Axelsson}, {Baldini}, {Ballet}, {Barbiellini}, {Bastieri},
  {Belfiore}, {Bellazzini}, {Bissaldi}, {Blandford}, {Bloom}, {Bogart},
  {Bonino}, {Bottacini}, {Bregeon}, {Britto}, {Bruel}, {Buehler}, {Burnett},
  {Buson}, {Caliand ro}, {Cameron}, {Caputo}, {Caragiulo}, {Caraveo},
  {Casandjian}, {Cavazzuti}, {Charles}, {Chaves}, {Chekhtman}, {Cheung},
  {Chiang}, {Chiaro}, {Ciprini}, {Claus}, {Cohen-Tanugi}, {Cominsky}, {Conrad},
  {Cutini}, {D'Ammando}, {de Angelis}, {DeKlotz}, {de Palma}, {Desiante},
  {Digel}, {Di Venere}, {Drell}, {Dubois}, {Dumora}, {Favuzzi}, {Fegan},
  {Ferrara}, {Finke}, {Franckowiak}, {Fukazawa}, {Funk}, {Fusco}, {Gargano},
  {Gasparrini}, {Giebels}, {Giglietto}, {Giommi}, {Giordano}, {Giroletti},
  {Glanzman}, {Godfrey}, {Grenier}, {Grondin}, {Grove}, {Guillemot}, {Guiriec},
  {Hadasch}, {Harding}, {Hays}, {Hewitt}, {Hill}, {Horan}, {Iafrate}, {Jogler},
  {J{\'o}hannesson}, {Johnson}, {Johnson}, {Johnson}, {Johnson}, {Kamae},
  {Kataoka}, {Katsuta}, {Kuss}, {La Mura}, {Land riu}, {Larsson}, {Latronico},
  {Lemoine-Goumard}, {Li}, {Li}, {Longo}, {Loparco}, {Lott}, {Lovellette},
  {Lubrano}, {Madejski}, {Massaro}, {Mayer}, {Mazziotta}, {McEnery},
  {Michelson}, {Mirabal}, {Mizuno}, {Moiseev}, {Mongelli}, {Monzani},
  {Morselli}, {Moskalenko}, {Murgia}, {Nuss}, {Ohno}, {Ohsugi}, {Omodei},
  {Orienti}, {Orlando}, {Ormes}, {Paneque}, {Panetta}, {Perkins},
  {Pesce-Rollins}, {Piron}, {Pivato}, {Porter}, {Racusin}, {Rando}, {Razzano},
  {Razzaque}, {Reimer}, {Reimer}, {Reposeur}, {Rochester}, {Romani},
  {Salvetti}, {S{\'a}nchez-Conde}, {Saz Parkinson}, {Schulz}, {Siskind},
  {Smith}, {Spada}, {Spandre}, {Spinelli}, {Stephens}, {Strong}, {Suson},
  {Takahashi}, {Takahashi}, {Tanaka}, {Thayer}, {Thayer}, {Thompson},
  {Tibaldo}, {Tibolla}, {Torres}, {Torresi}, {Tosti}, {Troja}, {Van Klaveren},
  {Vianello}, {Winer}, {Wood}, {Wood}, {Zimmer}, \& {Fermi-LAT
  Collaboration}}]{Acero_2015}
{Acero}, F., {Ackermann}, M., {Ajello}, M., {et~al.} 2015, \apjs, 218, 23

\bibitem[{{{Ackermann}} {et~al.}(2011){{Ackermann}}, {Ajello}, {Allafort},
  {Angelakis}, {Axelsson}, {Baldini}, {Ballet}, \&
  {Barbiellini}}]{Ackermann_2011}
{{Ackermann}}, M., {Ajello}, M., {Allafort}, A., {et~al.} 2011, \apj, 741, 30

\bibitem[{{Ackermann} {et~al.}(2011){Ackermann}, {Ajello}, {Allafort},
  {Angelakis}, {Axelsson}, {Baldini}, {Ballet}, {Barbiellini}, {Bastieri},
  {Bellazzini}, {Berenji}, {Blandford}, {Bloom}, {Bonamente}, {Borgland},
  {Bouvier}, {Bregeon}, {Brez}, {Brigida}, {Bruel}, {Buehler}, {Buson},
  {Caliandro}, {Cameron}, {Cannon}, {Caraveo}, {Casand jian}, {Cavazzuti},
  {Cecchi}, {Charles}, {Chekhtman}, {Cheung}, {Ciprini}, {Claus},
  {Cohen-Tanugi}, {Cutini}, {de Palma}, {Dermer}, {Silva}, {Drell}, {Dubois},
  {Dumora}, {Escande}, {Favuzzi}, {Fegan}, {Focke}, {Fortin}, {Frailis},
  {Fuhrmann}, {Fukazawa}, {Fusco}, {Gargano}, {Gasparrini}, {Gehrels},
  {Giglietto}, {Giommi}, {Giordano}, {Giroletti}, {Glanzman}, {Godfrey},
  {Grandi}, {Grenier}, {Guiriec}, {Hadasch}, {Hayashida}, {Hays}, {Healey},
  {J{\'o}hannesson}, {Johnson}, {Kamae}, {Katagiri}, {Kataoka},
  {Kn{\"o}dlseder}, {Kuss}, {Lande}, {Lee}, {Longo}, {Loparco}, {Lott},
  {Lovellette}, {Lubrano}, {Makeev}, {Max-Moerbeck}, {Mazziotta}, {McEnery},
  {Mehault}, {Michelson}, {Mizuno}, {Monte}, {Monzani}, {Morselli},
  {Moskalenko}, {Murgia}, {Naumann-Godo}, {Nishino}, {Nolan}, {Norris}, {Nuss},
  {Ohsugi}, {Okumura}, {Omodei}, {Orlando}, {Ormes}, {Ozaki}, {Paneque},
  {Pavlidou}, {Pelassa}, {Pepe}, {Pesce-Rollins}, {Pierbattista}, {Piron},
  {Porter}, {Rain{\`o}}, {Razzano}, {Readhead}, {Reimer}, {Reimer}, {Richards},
  {Romani}, {Sadrozinski}, {Scargle}, {Sgr{\`o}}, {Siskind}, {Smith},
  {Spandre}, {Spinelli}, {Strickman}, {Suson}, {Takahashi}, {Tanaka}, {Taylor},
  {Thayer}, {Thayer}, {Thompson}, {Torres}, {Tosti}, {Tramacere}, {Troja},
  {Vandenbroucke}, {Vianello}, {Vitale}, {Waite}, {Wang}, {Winer}, {Wood},
  {Yang}, \& {Ziegler}}]{radio_gamma_corr2}
{Ackermann}, M., {Ajello}, M., {Allafort}, A., {et~al.} 2011, \apj, 741, 30

\bibitem[{{Ackermann} {et~al.}(2017){Ackermann}, {Ajello}, {Baldini}, {Ballet},
  {Barbiellini}, {Bastieri}, {Becerra Gonzalez}, {Bellazzini}, {Bissaldi},
  {Blandford}, {Bloom}, {Bonino}, {Bottacini}, {Bregeon}, {Bruel}, {Buehler},
  {Buson}, {Cameron}, {Caragiulo}, {Caraveo}, {Cavazzuti}, {Cecchi}, {Cheung},
  {Chiang}, {Chiaro}, {Ciprini}, {Conrad}, {Costantin}, {Costanza}, {Cutini},
  {D'Ammando}, {de Palma}, {Desiante}, {Digel}, {Di Lalla}, {Di Mauro}, {Di
  Venere}, {Dom{\'\i}nguez}, {Drell}, {Favuzzi}, {Fegan}, {Ferrara}, {Finke},
  {Focke}, {Fukazawa}, {Funk}, {Fusco}, {Gargano}, {Gasparrini}, {Giglietto},
  {Giordano}, {Giroletti}, {Green}, {Grenier}, {Guillemot}, {Guiriec},
  {Hartmann}, {Hays}, {Horan}, {Jogler}, {J{\'o}hannesson}, {Johnson}, {Kuss},
  {La Mura}, {Larsson}, {Latronico}, {Li}, {Longo}, {Loparco}, {Lovellette},
  {Lubrano}, {Magill}, {Maldera}, {Manfreda}, {Marcotulli}, {Mazziotta},
  {Michelson}, {Mirabal}, {Mitthumsiri}, {Mizuno}, {Monzani}, {Morselli},
  {Moskalenko}, {Negro}, {Nuss}, {Ohsugi}, {Ojha}, {Omodei}, {Orienti},
  {Orlando}, {Ormes}, {Paliya}, {Paneque}, {Perkins}, {Persic},
  {Pesce-Rollins}, {Piron}, {Porter}, {Principe}, {Rain{\`o}}, {Rando}, {Rani},
  {Razzano}, {Razzaque}, {Reimer}, {Reimer}, {Romani}, {Sgr{\`o}}, {Simone},
  {Siskind}, {Spada}, {Spandre}, {Spinelli}, {Stalin}, {Stawarz}, {Suson},
  {Takahashi}, {Tanaka}, {Thayer}, {Thompson}, {Torres}, {Torresi}, {Tosti},
  {Troja}, {Vianello}, \& {Wood}}]{Ackermann_2017}
{Ackermann}, M., {Ajello}, M., {Baldini}, L., {et~al.} 2017, \apjl, 837, L5

\bibitem[{{Ajello} {et~al.}(2009){Ajello}, {Costamante}, {Sambruna}, {Gehrels},
  {Chiang}, {Rau}, {Escala}, {Greiner}, {Tueller}, {Wall}, \&
  {Mushotzky}}]{Ajello_2009}
{Ajello}, M., {Costamante}, L., {Sambruna}, R.~M., {et~al.} 2009, \apj, 699,
  603

\bibitem[{{Ajello} {et~al.}(2014){Ajello}, {Romani}, {Gasparrini}, {Shaw},
  {Bolmer}, {Cotter}, {Finke}, {Greiner}, {Healey}, {King}, {Max-Moerbeck},
  {Michelson}, {Potter}, {Rau}, {Readhead}, {Richards}, \&
  {Schady}}]{Ajello_2014}
{Ajello}, M., {Romani}, R.~W., {Gasparrini}, D., {et~al.} 2014, \apj, 780, 73

\bibitem[{{Angioni}(2018)}]{Angioni_2018}
{Angioni}, R. 2018, The Astronomer's Telegram, 11137

\bibitem[{{Bloemen} {et~al.}(1995){Bloemen}, {Bennett}, {Blom}, {Collmar},
  {Hermsen}, {Lichti}, {Morris}, {Schoenfelder}, {Stacy}, {Strong}, \&
  {Winkler}}]{Bloemen_1995}
{Bloemen}, H., {Bennett}, K., {Blom}, J.~J., {et~al.} 1995, \aap, 293, L1

\bibitem[{{Celotti} \& {Ghisellini}(2008)}]{Celotti_2008}
{Celotti}, A., \& {Ghisellini}, G. 2008, \mnras, 385, 283

\bibitem[{{Cheung}(2016)}]{Cheung_2016}
{Cheung}, C. 2016, The Astronomer's Telegram, 9854

\bibitem[{{{Cheung}}(2017)}]{Cheung_2017}
{{Cheung}}, C. 2017, The Astronomer's Telegram, 10356

\bibitem[{{Ciprini} \& {Thompson}(2013)}]{FA_paper}
{Ciprini}, S., \& {Thompson}, D.~J. 2013, arXiv e-prints, arXiv:1303.4054

\bibitem[{{Condon} {et~al.}(1998){Condon}, {Cotton}, {Greisen}, {Yin},
  {Perley}, {Taylor}, \& {Broderick}}]{Condon_1998}
{Condon}, J.~J., {Cotton}, W.~D., {Greisen}, E.~W., {et~al.} 1998, \aj, 115,
  1693

\bibitem[{Eitan \& Behar(2013)}]{flattening_xray}
Eitan, A., \& Behar, E. 2013, The Astrophysical Journal, 774, 29

\bibitem[{Fabian {et~al.}(2001)Fabian, Celotti, Iwasawa, \&
  Ghisellini}]{intrinsic_absorption}
Fabian, A.~C., Celotti, A., Iwasawa, K., \& Ghisellini, G. 2001, Monthly
  Notices of the Royal Astronomical Society, 324, 628

\bibitem[{{Fabian} {et~al.}(2001){Fabian}, {Celotti}, {Iwasawa}, {McMahon},
  {Carilli}, {Brandt}, {Ghisellini}, \& {Hook}}]{warm_absorber}
{Fabian}, A.~C., {Celotti}, A., {Iwasawa}, K., {et~al.} 2001, \mnras, 323, 373

\bibitem[{{Fan} {et~al.}(2016){Fan}, {Yang}, {Liu}, {Luo}, {Lin}, {Yuan},
  {Xiao}, {Zhou}, {Hua}, \& {Pei}}]{Fan_2016}
{Fan}, J.~H., {Yang}, J.~H., {Liu}, Y., {et~al.} 2016, \apjs, 226, 20

\bibitem[{{Feldman} \& {Cousins}(1998)}]{Feldman_1998}
{Feldman}, G.~J., \& {Cousins}, R.~D. 1998, \prd, 57, 3873

\bibitem[{{Foschini}(2009)}]{curved_EC_emission2}
{Foschini}, L. 2009, Advances in Space Research, 43, 1036

\bibitem[{{Fuhrmann} {et~al.}(2014){Fuhrmann}, {Larsson}, {Chiang},
  {Angelakis}, {Zensus}, {Nestoras}, {Krichbaum}, {Ungerechts}, {Sievers},
  {Pavlidou}, {Readhead}, {Max-Moerbeck}, \& {Pearson}}]{radio_gamma_corr}
{Fuhrmann}, L., {Larsson}, S., {Chiang}, J., {et~al.} 2014, \mnras, 441, 1899

\bibitem[{{Ghirlanda} {et~al.}(2010){Ghirlanda}, {Ghisellini}, {Tavecchio}, \&
  {Foschini}}]{Ghirlanda_2010}
{Ghirlanda}, G., {Ghisellini}, G., {Tavecchio}, F., \& {Foschini}, L. 2010,
  \mnras, 407, 791

\bibitem[{{Ghisellini}(2013)}]{Ghisellini_2013}
{Ghisellini}, G. 2013, \memsai, 84, 719

\bibitem[{{Ghisellini} {et~al.}(2010){Ghisellini}, {Della Ceca}, {Volonteri},
  {Ghirland a}, {Tavecchio}, {Foschini}, {Tagliaferri}, {Haardt}, {Pareschi},
  \& {Grindlay}}]{Ghisellini_2010b}
{Ghisellini}, G., {Della Ceca}, R., {Volonteri}, M., {et~al.} 2010, \mnras,
  405, 387

\bibitem[{{H.~E.~S.~S. Collaboration} {et~al.}(2013){H.~E.~S.~S.
  Collaboration}, {Abramowski}, {Acero}, {Aharonian}, {Akhperjanian}, {Anton},
  {Balenderan}, {Balzer}, {Barnacka}, {Becherini}, {Becker Tjus},
  {Bernl{\"o}hr}, {Birsin}, {Biteau}, {Bochow}, {Boisson}, {Bolmont}, {Bordas},
  {Brucker}, {Brun}, {Brun}, {Bulik}, {Carrigan}, {Casanova}, {Cerruti},
  {Chadwick}, {Charbonnier}, {Chaves}, {Cheesebrough}, {Cologna}, {Conrad},
  {Couturier}, {Dalton}, {Daniel}, {Davids}, {Degrange}, {Deil}, {deWilt},
  {Dickinson}, {Djannati-Ata{\"\i}}, {Domainko}, {O'C. Drury}, {Dubus},
  {Dutson}, {Dyks}, {Dyrda}, {Egberts}, {Eger}, {Espigat}, {Fallon}, {Farnier},
  {Fegan}, {Feinstein}, {Fernandes}, {Fernandez}, {Fiasson}, {Fontaine},
  {F{\"o}rster}, {F{\"u}{\ss}ling}, {Gajdus}, {Gallant}, {Garrigoux}, {Gast},
  {Giebels}, {Glicenstein}, {Gl{\"u}ck}, {G{\"o}ring}, {Grondin},
  {H{\"a}ffner}, {Hague}, {Hahn}, {Hampf}, {Harris}, {Heinz}, {Heinzelmann},
  {Henri}, {Hermann}, {Hillert}, {Hinton}, {Hofmann}, {Hofverberg}, {Holler},
  {Horns}, {Jacholkowska}, {Jahn}, {Jamrozy}, {Jung}, {Kastendieck},
  {Katarzy{\'n}ski}, {Katz}, {Kaufmann}, {Kh{\'e}lifi}, {Klochkov},
  {Klu{\'z}niak}, {Kneiske}, {Komin}, {Kosack}, {Kossakowski}, {Krayzel},
  {Laffon}, {Lamanna}, {Lenain}, {Lennarz}, {Lohse}, {Lopatin}, {Lu},
  {Marandon}, {Marcowith}, {Masbou}, {Maurin}, {Maxted}, {Mayer}, {McComb},
  {Medina}, {M{\'e}hault}, {Menzler}, {Moderski}, {Mohamed}, {Moulin},
  {Naumann}, {Naumann-Godo}, {de Naurois}, {Nedbal}, {Nguyen}, {Niemiec},
  {Nolan}, {Ohm}, {de O{\~n}a Wilhelmi}, {Opitz}, {Ostrowski}, {Oya}, {Panter},
  {Parsons}, {Paz Arribas}, {Pekeur}, {Pelletier}, {Perez}, {Petrucci},
  {Peyaud}, {Pita}, {P{\"u}hlhofer}, {Punch}, {Quirrenbach}, {Raue}, {Reimer},
  {Reimer}, {Renaud}, {de los Reyes}, {Rieger}, {Ripken}, {Rob}, {Rosier-Lees},
  {Rowell}, {Rudak}, {Rulten}, {Sahakian}, {Sanchez}, {Santangelo},
  {Schlickeiser}, {Schulz}, {Schwanke}, {Schwarzburg}, {Schwemmer}, {Sheidaei},
  {Skilton}, {Sol}, {Spengler}, {Stawarz}, {Steenkamp}, {Stegmann}, {Stinzing},
  {Stycz}, {Sushch}, {Szostek}, {Tavernet}, {Terrier}, {Tluczykont},
  {Valerius}, {van Eldik}, {Vasileiadis}, {Venter}, {Viana}, {Vincent},
  {V{\"o}lk}, {Volpe}, {Vorobiov}, {Vorster}, {Wagner}, {Ward}, {White},
  {Wierzcholska}, {Wouters}, {Zacharias}, {Zajczyk}, {Zdziarski}, {Zech}, \&
  {Zechlin}}]{Hess_2013}
{H.~E.~S.~S. Collaboration}, {Abramowski}, A., {Acero}, F., {et~al.} 2013,
  \aap, 550, A4

\bibitem[{{Houck} \& {Denicola}(2000)}]{Houck2000}
{Houck}, J.~C., \& {Denicola}, L.~A. 2000, {ISIS: An Interactive Spectral
  Interpretation System for High Resolution X-Ray Spectroscopy}

\bibitem[{{Li} {et~al.}(2020){Li}, {Sun}, {Liao}, \& {Fan}}]{Li_2020}
{Li}, S., {Sun}, L.-M., {Liao}, N.-H., \& {Fan}, Y.-Z. 2020, arXiv e-prints,
  arXiv:2006.10470

\bibitem[{{Liao} {et~al.}(2019){Liao}, {Dou}, {Jiang}, {Wang}, {Fan}, \&
  {Wang}}]{Liao_2019}
{Liao}, N.-H., {Dou}, L.-M., {Jiang}, N., {et~al.} 2019, \apjl, 879, L9

\bibitem[{{Liao} {et~al.}(2018){Liao}, {Li}, \& {Fan}}]{Liao_2018}
{Liao}, N.-H., {Li}, S., \& {Fan}, Y.-Z. 2018, \apjl, 865, L17

\bibitem[{{Marcotulli} {et~al.}(2020){Marcotulli}, {Paliya}, {Ajello}, {Kaur},
  {Marchesi}, {Rajagopal}, {Hartmann}, {Gasparrini}, {Ojha}, \&
  {Madejski}}]{Marcotulli_2020}
{Marcotulli}, L., {Paliya}, V., {Ajello}, M., {et~al.} 2020, arXiv e-prints,
  arXiv:2001.01956

\bibitem[{{Mattox} {et~al.}(1996){Mattox}, {Bertsch}, {Chiang}, {Dingus},
  {Digel}, {Esposito}, {Fierro}, {Hartman}, {Hunter}, {Kanbach}, {Kniffen},
  {Lin}, {Macomb}, {Mayer-Hasselwander}, {Michelson}, {von Montigny},
  {Mukherjee}, {Nolan}, {Ramanamurthy}, {Schneid}, {Sreekumar}, {Thompson}, \&
  {Willis}}]{Mattox_1996}
{Mattox}, J.~R., {Bertsch}, D.~L., {Chiang}, J., {et~al.} 1996, \apj, 461, 396

\bibitem[{{Mereu} {et~al.}(in prep){Mereu}, {Cavazzuti}, {Cutini}, \&
  {Tosti}}]{Mereu_inprep}
{Mereu}, I., {Cavazzuti}, E., {Cutini}, S., \& {Tosti}, G. in prep, \aap

\bibitem[{{Orienti} {et~al.}(2014){Orienti}, {D'Ammando}, {Giroletti}, {Finke},
  {Ajello}, {Dallacasa}, \& {Venturi}}]{Orienti_2014}
{Orienti}, M., {D'Ammando}, F., {Giroletti}, M., {et~al.} 2014, \mnras, 444,
  3040

\bibitem[{{Paliya} {et~al.}(2020){Paliya}, {Ajello}, {Cao}, {Giroletti},
  {Kaur}, {Madejski}, {Lott}, \& {Hartmann}}]{Paliya_2020}
{Paliya}, V.~S., {Ajello}, M., {Cao}, H.~M., {et~al.} 2020, \apj, 897, 177

\bibitem[{{Paliya} {et~al.}(2016){Paliya}, {Parker}, {Fabian}, \&
  {Stalin}}]{Paliya_2016}
{Paliya}, V.~S., {Parker}, M.~L., {Fabian}, A.~C., \& {Stalin}, C.~S. 2016,
  \apj, 825, 74

\bibitem[{{Paliya} {et~al.}(2019{\natexlab{a}}){Paliya}, {Ajello}, {Ojha},
  {Angioni}, {Cheung}, {Tanada}, {Pursimo}, {Galindo}, {Losada}, {Siltala},
  {Djupvik}, {Marcotulli}, \& {Hartmann}}]{Paliya_2019}
{Paliya}, V.~S., {Ajello}, M., {Ojha}, R., {et~al.} 2019{\natexlab{a}}, \apj,
  871, 211

\bibitem[{{Paliya} {et~al.}(2019{\natexlab{b}}){Paliya}, {Ajello},
  {Marcotulli}, {Tomsick}, {Perkins}, {Prandini}, {D'Ammando}, {De Angelis},
  {Thompson}, {Li}, {Dominguez}, {Beckmann}, {Guiriec}, {Wadiasingh}, {Coppi},
  {Harding}, {Petropoulou}, {Hewitt}, {Ojha}, {Marcowith}, {Doro}, {Castro},
  {Baring}, {Hays}, {Orlando}, {Guiriec}, {Bozhilov}, {Agudo}, {Venters},
  {McEnery}, {The}, {Hartmann}, {Buson}, {Longo}, \&
  {Gasparrini}}]{Paliya_2019b}
{Paliya}, V.~S., {Ajello}, M., {Marcotulli}, L., {et~al.} 2019{\natexlab{b}},
  arXiv e-prints, arXiv:1903.06106

\bibitem[{Tavecchio {et~al.}(2007)Tavecchio, Maraschi, Ghisellini, Kataoka,
  Foschini, Sambruna, \& Tagliaferri}]{curved_EC_emission}
Tavecchio, F., Maraschi, L., Ghisellini, G., {et~al.} 2007, The Astrophysical
  Journal, 665, 980

\bibitem[{{Toda} {et~al.}(2020){Toda}, {Fukazawa}, \& {Inoue}}]{Toda_2020}
{Toda}, K., {Fukazawa}, Y., \& {Inoue}, Y. 2020, arXiv e-prints,
  arXiv:2005.02648

\bibitem[{{van den Berg} {et~al.}(2019){van den Berg}, {B{\"o}ttcher},
  {Dom{\'\i}nguez}, \& {L{\'o}pez-Moya}}]{Berg_2019}
{van den Berg}, J.~P., {B{\"o}ttcher}, M., {Dom{\'\i}nguez}, A., \&
  {L{\'o}pez-Moya}, M. 2019, \apj, 874, 47

\bibitem[{Wilks(1938)}]{Wilks_1938}
Wilks, S.~S. 1938, Annals Math. Statist., 9, 60

\bibitem[{{Yuan} {et~al.}(2000){Yuan}, {Matsuoka}, {Wang}, {Ueno}, {Kubo}, \&
  {Mihara}}]{intrinsic_absorption2}
{Yuan}, W., {Matsuoka}, M., {Wang}, T., {et~al.} 2000, \apj, 545, 625

\end{thebibliography}

 \begin{table*}[!pt]
  ~\vspace{-25pt}
     \caption{The 27 high-$z$ blazars that were identified, each having been detected
  with TS\,$\geq 9$ in more than one monthly interval. The sources listed
  in Table \,\ref{High-Z sources subsample} as well as mis-identified sources and blazars that were listed in 3FGL have been excluded. Out of this sample, 23 sources do not show any known $\gamma$-ray counterpart. Significantly identified previously undetected high-\textit{z} blazars are marked in bold.}
 \centering
 \begin{footnotesize}
     \begin{tabular}{c c c c c c c}
     \label{High-Z sources TS 9}
 \begin{tabular}{c} \textbf{Source} \\  \textbf{Name}     \end{tabular}    &    \begin{tabular}{c} \textbf{RA} \\  \textbf{J2000}     \end{tabular} &  \begin{tabular}{c} \textbf{DEC} \\  \textbf{J2000}     \end{tabular}  & \textbf{z}  & \begin{tabular}{c} \textbf{Detections} \\  \textbf{TS\,$\geq \text{9}$}     \end{tabular} & \begin{tabular}{c} \textbf{$\text{P}_{\text{tot}}$} \\  \textbf{$[\sigma]$}     \end{tabular}  & \begin{tabular}{c} \textbf{$\gamma$-ray} \\ \textbf{Counterpart} \end{tabular}\\\\\hline
 5BZQ\,J1429+5406 	 &217.34	 &54.10	 &3.01	 &6	 &7.75	 &4FGL J1428.9+5406\\
 \textbf{5BZQ\,J2321$-$0827}  &\textbf{350.33} & \textbf{-8.46}  &\textbf{3.16}	 &\textbf{5}	 &\textbf{6.84}	 &\textbf{-/-}\\
\textbf{5BZQ\,J1430+4204}	 & \textbf{217.60}  & \textbf{42.08}	 & \textbf{4.72}	 & \textbf{5}  &\textbf{6.57}	  & \textbf{-/-}\\
 \textbf{5BZU\,J0525$-$3343}   & \textbf{81.28}	 & \textbf{-33.72} & \textbf{4.41}  &\textbf{4}  & \textbf{5.73}  & \textbf{-/-}\\
  5BZQ\,J1124$-$1501  & 171.01	 & -15.03	 & 2.56	 & 3	 & 5.50	 & 4FGL J1122.5$-$1440 \\
  \textbf{5BZQ\,J0741+2557} 	 & \textbf{115.37}	 & \textbf{25.96}	 & \textbf{2.75}	 & \textbf{3}	 & \textbf{5.49} & 	\textbf{-/-}\\
  \textbf{5BZQ\,J1424+2256} &	\textbf{216.16} &	\textbf{22.93} &	\textbf{3.62} &	\textbf{4} &	\textbf{5.25} &	\textbf{-/-}\\
  \textbf{5BZQ\,J1007+1356} 	 &\textbf{151.92} &	\textbf{13.94} &	\textbf{2.72} &	\textbf{3} &	\textbf{5.24} &	\textbf{-/-}\\
  5BZQ\,J0220$-$2151 &  	35.15 &	-21.85 &	2.55 &	2 &	4.53 &	-/-\\
  5BZQ\,J0530$-$2503 & 	82.53 &	-25.06 &	2.78 &	2 &	4.52 &	4FGL J0529.1$-$2449\\
 5BZQ\,J0839+5112 & 	129.94 &	51.20 &	4.40 &	2 &	4.52 &	-/-\\
  5BZQ\,J1949$-$1957 & 	297.47 &	-19.95 &	2.65 &	2 &	4.52 &	-/-\\
 5BZQ\,J1740+4506 &	265.03 &	45.11 &	2.79 &	2 &	4.51 &	-/-\\
 5BZQ\,J1942$-$7015 & 	295.69 &	-70.26 &	2.90 &	2 &	4.50 &	-/-\\
  5BZQ\,J0324$-$2918 &	51.18 &	-29.31 &	4.63 &	2 &	4.49 &	-/-\\
 5BZU\,J2248$-$0541 & 	342.00 &	-5.69 &	3.30 &	2 &	4.48 &	-/-\\
 5BZQ\,J1616+0459 &	244.16 &	4.99 &	3.22 &	2 &	4.48 &	-/-\\
 5BZQ\,J0121$-$2806 & 	20.25 &	-28.11 &	3.12 &	2 &	4.48 &	-/-\\
 5BZQ\,J1413+4505 & 	213.33 &	45.09 &	3.11 &	2 &	4.48 &	-/-\\
 5BZQ\,J1421$-$0643 & 	215.28 &	-6.73 &	3.69 &	2 &	4.48 &	-/-\\
 5BZQ\,J1100$-$4249 & 	165.21 &	-42.82 &	3.56 &	3 &	4.47 &	-/-\\
 SHAO\,J0257+4338 & 	44.50 &	43.64 &	4.07 &	2 &	4.47 &	-/-\\
 5BZQ\,J2015+6554 & 	303.98 &	65.91 &	2.85 &	2 &	4.47 &	4FGL J2015.4+6556\\
 5BZQ\,J0001+1914 & 	0.29 &	19.24 &	3.10 &	2 &	4.46 &	-/-\\
 5BZQ\,J0612$-$4337 & 	93.12 &	-43.63 &	3.46 &	3 &	4.46 &-/-\\	
 5BZQ\,J0918+0636 & 	139.60 &	6.61 &	4.22 &	2 &	4.45 &	-/-\\
  5BZQ\,J1139$-$1552  & 	174.87	 &-15.88 &	2.63 &	2 &	3.53 &	-/-\\
  \\\hline
\end{tabular}
 \end{footnotesize}
  \end{table*}

\end{document}